\documentclass[aps, longbibliography, superscriptaddress]{revtex4-2}
\usepackage{amsmath}
\usepackage{graphicx, textcomp}
\usepackage{enumerate,subeqnarray}
\usepackage{verbatim,natbib}
\usepackage{amsmath, amsthm, amssymb}
\usepackage{amsfonts}
\usepackage{subcaption, caption}
\usepackage[colorlinks=true, citecolor=red, linkcolor=blue, urlcolor=blue ]{hyperref}
\usepackage{booktabs, multirow}

\usepackage{orcidlink, etoolbox}

\begin{document}
\title{Fluid mechanics of sarcomeres as porous media}
\author{John Severn\orcidlink{0009-0001-2098-4997}}
\affiliation{Department of Applied Mathematics and Theoretical Physics, University of Cambridge,	
		Cambridge CB3 0WA, UK}
\author{Thomas Vacus\orcidlink{0009-0006-6976-9725}}
\affiliation{Department of Applied Mathematics and Theoretical Physics, University of Cambridge,	
		Cambridge CB3 0WA, UK}
\affiliation{Laboratoire de Physique de l'Ecole Normale Supérieure, Université PSL, 75005 Paris, France}
\author{Eric Lauga\orcidlink{0000-0002-8916-2545}}
\email{e.lauga@damtp.cam.ac.uk}
\affiliation{Department of Applied Mathematics and Theoretical Physics, University of Cambridge,	
		Cambridge CB3 0WA, UK}
	\date{\today}
	\begin{abstract}

Muscle contraction, both in skeletal and cardiac tissue, is driven by sarcomeres, the microscopic units inside muscle cells where thick myosin and thin actin filaments slide past each other. During contraction and relaxation, the sarcomere's volume changes, causing sarcoplasm (intra-sarcomeric fluid) to flow out during contraction and back in as the sarcomere relaxes. We present a quantitative model of this sarcoplasmic flow, treating the sarcomere as an anisotropic porous medium with regions defined by the presence and absence of thick and thin filaments. Using semi-analytic methods, we solve for axial and lateral fluid flow within the filament lattice, calculating the permeabilities of this porous structure. 
We then apply these permeabilities within a Darcy model to determine the flow field generated during contraction. The predictions of our continuum model show excellent agreement with finite element simulations, reducing computational time by several orders of magnitude while maintaining  accuracy in modelling the biophysical flow dynamics.

	\end{abstract}
	\maketitle

\section{Introduction}

\begin{figure}
\centering
\includegraphics[width = 0.4\textwidth]{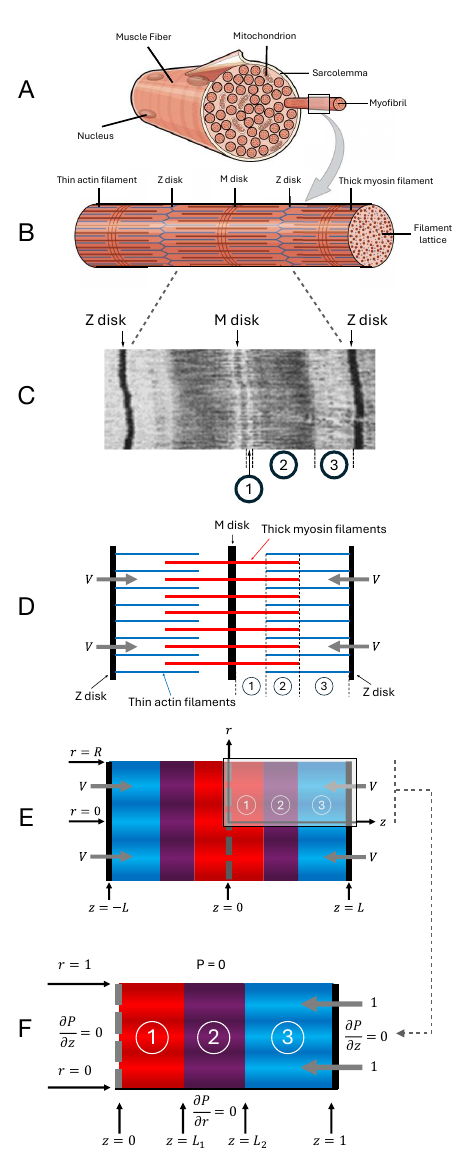}
\caption{Physical and mathematical modelling of sarcomere. 
(\textbf{A}) A muscle fibre containing many myofibrils~\cite{OpenStax_Edited}.
{(\textbf{B}) A single myofibril, containing many aligned sarcomeres. Each Z disk defines the boundary from one sarcomere to the next~\cite{OpenStax_Edited}.}
(\textbf{C}) Electron micrograph of a single sarcomere, showing the defining banding patterns of striated muscle~\cite{Sameerb}; regions of different intensities correspond to distinct regions in the sarcomere{, which are labelled as in D}.
(\textbf{D}) Simplified structure of a single sarcomere, with thin actin filaments (blue), thick myosin filaments (red), and Z and M disks;  the three regions are defined by the presence of the thick and/or thin filaments:  regions 1 (thick only),  2 (overlap) and  3 (thin only).
(\textbf{E}) Darcy model representation of the sarcomere as an anisotropic porous medium (with the three regions highlighted in different colours), showing radius $R$ and overall sarcomere length $2L$. 
(\textbf{F}) Blow-up of top-right quarter of the Darcy model representation, with non-dimensionalisations giving radius and half-length of $1$, and boundary conditions as indicated.}
\label{Figure 1}
\end{figure}

Many biological processes, ranging from locomotion of microorganisms to large-scale muscle movement, rely on molecular motors, nano-scale machines that    generate motion~\cite{Alberts_Wilson_Hunt_Heald_Johnson_Morgan_Raff_Roberts_Walter_2022}. One such molecular motor is the actin-myosin motor, which converts energy stored in ATP molecules into movement when myosin filaments repeatedly bind (also known as cross-bridging) and unbind to parallel actin filaments. This process is observed as a characteristic ``walking" motion of the myosin along the actin, and the two filaments slide relative to one another~\cite{Lymn_Taylor_1971, Al-Khayat_2013}. One  {notable} example of biological components that utilise the actin-myosin motor are sarcomeres,  the fundamental contractile units of so-called striated  muscle (both skeletal and cardiac), composed of periodic hexagonal arrays of axially-aligned, interdigitating (thin) actin and (thick)  myosin filaments~\cite{Huxley_1953, Huxley_1957}.  {Sarcomeres are approximately cylindrical, with a typical diameter of one micron and a typical length of a few microns~\cite{Powers_Nishikawa_Joumaa_Herzog_2016, Hou_2018, González-Morales_Xiao_Schilling_Marescal_Liao_Schöck_2019, Shimomura_Iwamoto_Doan_Ishiwata_Sato_Suzuki_2016}.} Many micro-scale sarcomeres arranged end-to-end form a myofibril, and it is the collective periodic contraction and relaxation of many myofibrils that produce large-scale muscle movement (see illustrations in Fig.~\ref{Figure 1}A {and \ref{Figure 1}B}).  {In this paper, we address the fluid mechanics of sarcomere contraction; we first present in the Introduction a broad overview of the biophysics of sarcomere contraction before detailing the motivation for our modelling study.} 

There is great variation in the precise structure of sarcomeres across different species, and even a single organism can contain a range of different sarcomeres~\cite{Craig_Padron_2004, Wray_Vibert_Cohen_1975, Page_1965, Peachey_Huxley_1962, Shimomura_Iwamoto_Doan_Ishiwata_Sato_Suzuki_2016}. However, all sarcomeres of striated muscles have the same characteristic banding pattern (see electron micrograph in Fig.~\ref{Figure 1}C): an interdigitating array of thick and thin filaments attached to Z disks (Fig.~\ref{Figure 1}D). 
The Z disks are thick and dense structures of interconnecting proteins that separate adjacent sarcomeres~\cite{Luther_2000, Luther_Barry_Squire_2002, Schroeter_Bretaudiere_Sass_Goldstein_1996}. Each Z disk anchors an array of thin filaments, composed primarily of actin~\cite{Holmes_Popp_Gebhard_Kabsch_1990, Pollard_1990}. Each sarcomere also contains an array of thick filaments, composed mostly of myosin~\cite{Huxley_1963, Trinick_Elliott_1979, Knight_Trinick_1984, Huxley_Brown_1967, Kensler_Stewart_1983}. The midpoints of the thick filaments are anchored to the M disk, which is a thick, multilayered, dense structure of proteins that separates the two halves of the sarcomere~\cite{Knappeis_Carlsen_1968, Luther_Squire_1978, Luther_Crowther_1984, Crowther_Luther_1984} (this M disk is absent from some sarcomeres,  mostly in invertebrates~\cite{Page_1965, Peachey_Huxley_1962, Craig_Padron_2004}); as we will show below, the presence of M disks in fact has little impact on the fluid flow. We also note that thick filaments continue unbroken through the M disk (if present), with `half-filaments' on either side of the M disk; in what follows, we use the term thick half-filament to avoid confusion. 
Meanwhile, the thin filaments from one sarcomere do not continue unbroken through the Z disk, into the next sarcomere, and so the Z disk has entire thin filaments on either side. An example of the periodic hexagonal arrangement of filaments is illustrated in Fig.~\ref{Figure 2}A, which shows a cross-section of the overlap region of the sarcomere. Upon activation of the muscle, calcium ions are released into the sarcomere from the sarcoplasmic reticulum, {an intricate tubular structure that surrounds myofibrils \cite{Sommer_1982},} triggering myosin-actin cross-bridging~\cite{Zot_Potter_1987, Solaro_Rarick_1998, Gordon_Homsher_Regnier_2000} and causing the thin filaments (and the Z disks to which they are attached) to be pulled inwards with some speed $V${, on the order of $1~\mu$m/s~\cite{ter_Keurs_Diao_Deis_2010, Rodriguez_Hunter_Royce_Leppo_Douglas_Weisman_1992, Shankar_Mahadevan_2024a}}. This shortens the sarcomere, seen as contraction of the muscle~\cite{HUXLEY_NIEDERGERKE_1954, HUXLEY_HANSON_1954, Page_Huxley_1963, Huxley_Brown_1967, Elliott_Lowy_Millman_1967, Huxley_Faruqi_Kress_Bordas_Koch_1982}.

Despite the many  {protein filaments (i.e.~myofilaments)} within the sarcomere, the majority of the space between the M disk and Z disk is filled with the liquid cytoplasm of muscle cells, termed sarcoplasm in this context~\cite{TOUMANIDOU2018141}. {Studies of sarcomeres from various species have shown that inter-filament spacing and sarcomere radius can change slightly over the course of a contraction. However, radial strain proves insufficient to maintain a constant volume~\cite{Shankar_Mahadevan_2024a}. Consequently,  {assuming incompressibility,} this implies a  flow of sarcoplasm out of the sarcomere during contraction, and back in during relaxation. It has recently been proposed, and supported by initial investigation, that this fluid flow could augment the transport of substrates  such as calcium ions and ATP necessary for sarcomere function~\cite{Cass_Williams_Irving_Lauga_Malingen_Daniel_Sponberg_2021}.

The many myofilaments of the sarcomere can be modelled as long, slender cylinders, and provide complicated but periodic obstructions to fluid flow. The explicit flow past these filaments can be calculated numerically, such as by  {the} finite element method~\cite{Malingen_Hood_Lauga_Hosoi_Daniel_2021}. However, this not only involves designing an appropriate explicit model, but also requires specialised computational modelling or software, and significant computational power and time to produce accurate results. Furthermore, the process must be repeated whenever considering a different type of sarcomere, or the same sarcomere at a different stage of contraction.

In this paper, we propose an alternative modelling approach and derive a   theoretical model of fluid flow in and out of the simplified sarcomere of Fig.~\ref{Figure 1}D that circumvents the need to consider the  {myofilaments} individually and the concomitant small-scale fluid flow. The model instead considers the averaged large-scale fluid behaviour over many filaments, treating the sarcomere as an anisotropic porous medium, as summarised in Fig.~\ref{Figure 1}E. The results of this model lead to excellent agreement with full numerical computations that resolve  {all} the sarcomere length scales, but with vastly reduced computational cost, time and complexity.

 {This paper is split into two main sections, one devoted to the physical and mathematical modelling (\S\ref{sec:modelling}), and a second one discussing the results for flows inside sarcomeres and comparing our model with full numerical simulations (\S\ref{sec:results}). In \S\ref{sec:modelling}, we {justify some of our physical assumptions, and then} set up the geometry of the model and non-dimensionalise the problem. We next derive and solve the Darcy equations for the pressure and fluid flow in an anisotropic porous medium in each of the three regions of the sarcomere, subject to the relevant   boundary and interface conditions. These solutions involve numerical physical parameters {(namely, the permeabilities and the traction parameter)}, whose values we determine by calculating the small-scale fluid flow through the periodic cells of the filament lattice, via semi-analytic methods. We next  {compute} in \S\ref{sec:results} the large-scale fluid flow  {of the Darcy model,} and compare with results of full numerical computations,  {finding excellent agreement in several orders of magnitude less time. We then compare results of the Darcy model between different sarcomeres, identifying common flow properties, as well as differences. We conclude with a discussion (including potential improvements) of the modelling approach{, as well as potential applications,} in \S\ref{sec:discussion}.}

\section{Modelling flow within a sarcomere}\label{sec:modelling}

{\subsection{Physical assumptions and simplifications}
Our model is derived using  two simplifying  physical assumptions. We first model the sarcoplasm as a Newtonian fluid. Previous observations indicate that the cytoplasm of muscle fibres is approximately Newtonian for all but the fastest and highest frequency contractions \cite{de_Tombe_ter_Keurs_1992, Tanner_Farman_Irving_Maughan_Palmer_Miller_2012, Swank_Kronert_Bernstein_Maughan_2004, Shankar_Mahadevan_2024a}. However, since the non-Newtonian behaviour of cytoplasm is primarily due to the cytoskeleton \cite{Thekkethil_Kory_Guo_Stewart_Hill_Luo_2024, Najafi_2023}, which is absent from the interior of the sarcomere, the behaviour of sarcoplasm likely remains Newtonian even at such extremes. Nano-scale objects in the sarcoplasm, such as enzymes \cite{Hargreaves_Spriet_2020}, are likely insufficient, both in abundance \cite{Luby-Phelps_1999} and individual size, to cause non-Newtonian behaviour \cite{van_der_Werff_de_Kruif_Blom_Mellema_1989, Dhont_van_der_Werff_de_Kruif_1989, Verberg_de_Schepper_Cohen_1997}. 

Secondly, in the model we take the sarcomere radius (and the  inter-filament spacing) to be constant. While it is well established that sarcomere radius (and consequently filament spacing) is not constant throughout contraction,   radial strains (i.e. fractional changes in cross-sectional area) are generally smaller than axial strains (i.e. fractional changes in sarcomere length)~\cite{Shankar_Mahadevan_2024a}. The quadratic dependence of cross-sectional area on radius, and the fact that sarcomere radius is typically smaller than length (see Table~\ref{table1}, \cite{Kono_Kawai_Shimamoto_Ishiwata_2020, Washio_Shintani_Higuchi_Sugiura_Hisada_2019, Powers_Nishikawa_Joumaa_Herzog_2016}), indicate that absolute changes in radius are much smaller than absolute changes in length. Furthermore, the radial strain profile over the course of a contraction varies between species, and the Poisson ratio {(i.e.~the ratio of radial expansion to axial compression)} can vary between positive and negative values~\cite{Shankar_Mahadevan_2024a}, making precise and general modelling difficult. In terms of fluid flow (see \S\ref{sec:Darcy}), radial dilation of a matrix of fibres cannot drive any net fluid flux due to overall volume conservation, and we found that the changes to the permeability values (and especially their ratios) and the traction parameter are negligible, at most a few percent, for biologically observed radial strains \cite{Shankar_Mahadevan_2024a}. Therefore, it is appropriate to assume that the sarcomere radius, and inter-filament spacing, remain constant. In other words, we model the filaments as being rigid, not expecting this simplification to have any significant effect on fluid flow.}}

\subsection{Setup and non-dimensionalisation}\label{sec:setup}

{Following these assumptions,} we aim to solve for the flow inside the model sarcomere illustrated in Fig.~\ref{Figure 1}E. Our analysis, unless stated otherwise, will assume a dimensionless sarcomere  {described by cylindrical polar coordinates $(r, z, \theta)$, denoting the radial, axial and azimuthal coordinates respectively,} where axial  {($z$)} lengths are scaled with the sarcomere half-length $L$, radial  {($r$)} lengths with the sarcomere radius $R$ and axial velocities with the contraction speed $V$. As a result, the dimensionless sarcomere half-length, radius and contraction speed are all equal to $1$. The standard incompressibility equation $\nabla\cdot\mathbf{u}=0$ implies a scaling for lateral velocities of $RV/L$. Further, pressure is scaled with the viscous scaling $\mu V/L$, where $\mu$ is the dynamic viscosity of the fluid. We may finally exploit $z \to -z$ symmetry to solve the problem on a half-sarcomere.  {In this dimensionless problem, region 1 (thick filaments only) is located in $0 < z < L_1$, region 2 (overlap region) is located in $L_1 < z < L_2$, and region 3 (thin filaments only) is located in $L_2 < z < 1$.} 

We will show in our analysis that the porous medium exhibits lateral isotropy  {(all flow in a cross-section has the same permeability, regardless of direction)} which, coupled with the cylindrical geometry of the sarcomere, leads to azimuthal symmetry, and thus  we consider the problem in the $(r, z)$ polar plane. The resulting dimensionless setup is shown in Fig.~\ref{Figure 1}F (which also shows boundary conditions, that we discuss below).  Note that the three regions of the sarcomere (separated by interfaces at $z = L_1$ and $z = L_2$) are considered separately: we solve for the pressure and fluid flow in each region individually, and apply interface conditions to relate them. 

\subsection{Darcy flow modelling}\label{sec:Darcy}

\subsubsection{Introduction to Darcy flow.}

Within sarcomeres, there is a clear  separation of length scales between, on one side, the thick and thin filaments (typical width $10$~nm) and the spacing between the various filaments  (typical size $10$'s of nm) and, on the other, the sarcomere itself (typical width and length of 1~$\mu$m). Motivated by this observation, we model in our paper the sarcomere as a porous medium, i.e.~a medium that is not simply free fluid, but rather consists of fluid (occupying a volume fraction $\Phi$, termed the porosity) that must navigate fixed physical obstructions (here, the myofilaments, occupying a volume fraction $1 - \Phi$). The fluid/obstruction mixture is considered a continuum, which is valid assuming that the length scale of flow is much larger than the pore scale (the characteristic size and separation of the obstructions), as justified above. The resultant flow problem is known as Darcy flow, a statistical average of fluid flow over many pore scales~\cite{Bear_2018, Guyon_2015}. 

In our sarcomere model, we deviate from traditional Darcy flow in two ways. First, Darcy flow typically considers isotropic porous media, but here, the alignment of the myofilaments along the long axis of the sarcomere makes the medium anisotropic, with axial flow typically easier than lateral flow. Second, most porous media consist of pores that exhibit randomness, with their averaged properties determined statistically. Here, instead, the porous medium consists of regular periodic cells, and we perform averages over individual cells rather than many pore scales. Darcy flow is concerned not with the fluid velocity, but rather the fluid flux (often called the Darcy flux or Darcy velocity), which is the rate of fluid flux per unit area through the porous medium. Pressure also exists as a spatial average. To avoid confusion, we thus reserve throughout the terms $\mathbf{u}$ and $p$ for conventional, interstitial (fluid) velocity and pressure, respectively, which undergo sharp changes  {within} a periodic cell, and $\mathbf{U}$ and $P$ for the Darcy flux and Darcy pressure, which eliminate these sharp changes via spatial averaging, with $\mathbf{U} = \Phi\left<\mathbf{u}\right>$ and $P = \left<p\right>$.

\subsubsection{Darcy fluxes in  anisotropic porous medium.}

\begin{figure}[t]
\centering
\includegraphics[width = 0.47\textwidth]{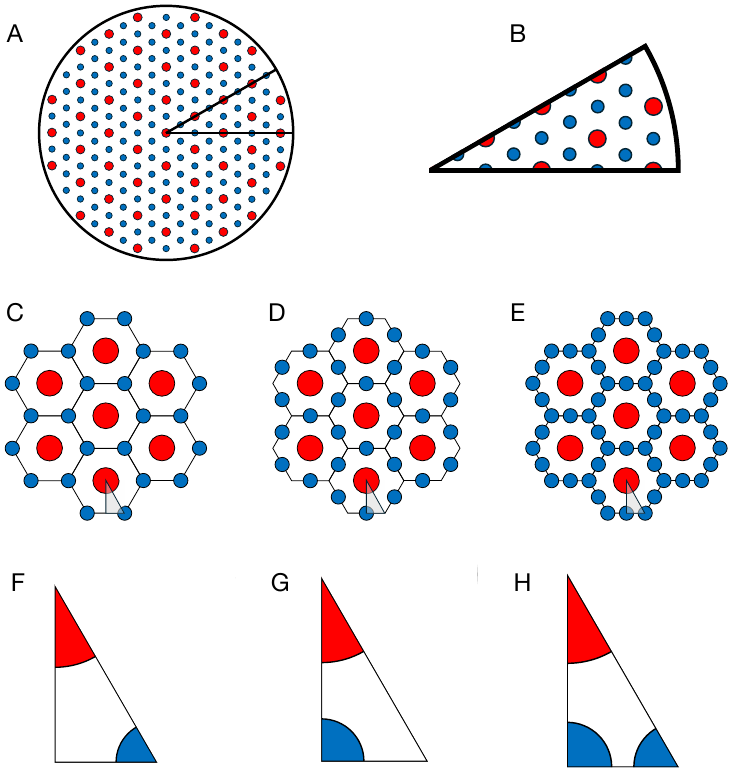}
\caption{(\textbf{A}) Sarcomere cross-section in the overlap region (i.e.~region 2 in Fig.~\ref{Figure 1}D), here illustrated for a sarcomere of hypothetical radius $200~$nm{, with a $30^\circ$ sector of symmetry indicated}. (\textbf{B}) $30^\circ$ sector of cross-section{, as indicated in A}. (\textbf{C}-\textbf{E}) Hexagonal arrangement of filaments in the overlap region. Packing ratios (number of thin filaments per thick half-filament) of 2 (C),  3 (D) and 5  (E) are shown, with 2 being typical for vertebrates and 3 and 5 often being found in invertebrates~\cite{Stenger_Spiro_1961, HOYLE_1967, Shimomura_Iwamoto_Doan_Ishiwata_Sato_Suzuki_2016}. {Fundamental triangles, each being $1/12$ of a hexagon, are shown at the bottom. (\textbf{F}-\textbf{H}) Fundamental triangles, highlighted in C-E, used to compute the permeabilities and the traction parameter $\beta$.}}
\label{Figure 2}
\end{figure}

In order to  {identify the} large-scale flow behaviour in the sarcomere, we first consider the small-scale flow between the filaments and properly derive the Darcy model. To do so, we focus on the individual periodic hexagonal cells that comprise the larger periodic lattice, as shown in Fig.~\ref{Figure 2}C-E. By neglecting variation in the flow between adjacent periodic cells, we exploit various symmetries to  {calculate} semi-analytic expressions for the fluid flow within a single periodic cell, and obtain  explicit linear relationships between fluid flux and pressure gradient. Details of the derivations, including necessary assumptions, are given in Appendix~\ref{appendix:axial_permeabilities} (axial permeabilities), \ref{appendix:beta} (traction parameter $\beta$), \ref{appendix:lateral_permeabilities} (lateral permeabilities), but essentially we need only assume that the length scales (both axial and lateral) over which $\nabla P$ varies are much larger than the size of each individual hexagonal cell of myofilaments. 

By solving for the axial fluid flow in a hexagonal cell, we calculate an overall axial fluid flux per unit area of 
\begin{equation}
U_z = -k_\|\frac{\partial P}{\partial z}, 
\end{equation} for some constant axial permeability $k_\|$ that is a function only of the geometry of the hexagonal lattice and  which of the three regions we are in (thin filaments vs thick filaments vs overlap region); the various axial permeabilities are calculated in Appendix~\ref{appendix:axial_permeabilities}.

We now turn our attention to the lateral fluid flow, that is, fluid flow within a cross-section of the cylindrical sarcomere, perpendicular to the $z$ axis. In general, the resultant flux can be expressed in terms of a $2 \times 2$ lateral permeability matrix $\mathbf{k}_\perp$, where $\perp$ indicates the plane perpendicular to the $z$ axis:
\begin{equation}
\mathbf{U}_\perp = -\mathbf{k}_\perp\nabla_\perp P,
\end{equation}
where $\nabla_\perp P$ denotes the 2-dimensional gradient of $P$ in the $(r, \theta)$ plane, perpendicular to the $z$ axis. Once again, we calculate $\mathbf{k}_\perp$ by deriving the explicit lateral fluid flow within a hexagonal cell. Note, however, that the hexagonal cells, in Fig.~\ref{Figure 2}C-E, exhibit a six-fold rotational symmetry. In particular, this means that a lateral pressure gradient $\nabla_\perp P$ (of fixed magnitude) in the direction normal to any given side of the hexagons will yield the same magnitude of lateral fluid flux per unit area, i.e.~Darcy flux, and this flux will be directed in the same direction as the applied pressure gradient. Importantly, two such (independent) directions form a basis of the plane. We further note that the linearity of the Stokes equations allows us to reconstruct any flow (including the corresponding pressure field) as a linear combination of said basis. We consequently deduce that the lateral Darcy flux is equal in magnitude, and parallel to $\nabla_\perp P$, regardless of the direction of the applied pressure gradient. We therefore calculate a lateral flux per unit area $\mathbf{U}_{\perp} = -k_\perp\nabla_\perp P$ for some constant scalar lateral permeability $k_\perp$ (here also only a function of local lattice geometry and which of the three regions of interest is being considered), which we calculate in Appendix~\ref{appendix:lateral_permeabilities}. 

We conclude that the Darcy model of the sarcomere has cross-sectional isotropy, at which point the {cylindrical} geometry of the entire porous sarcomere implies an overall Darcy fluid flow that is independent of the azimuthal angle $\theta$ and has no component in the $\theta$ direction, but is instead purely axial and radial. We therefore obtain the  Darcy equations for flow in a sarcomere as a porous medium as
\begin{equation}\label{eq:Darcy1}
U_z^{(j)} = -k_{\|}^{(j)}\frac{\partial P^{(j)}}{\partial z} ,  \qquad U_r^{(j)} = -k_{\perp}^{(j)}\frac{\partial P^{(j)}}{\partial r},
\end{equation}
where we have used  superscript $j$ to refer to the three regions of the sarcomere, and no Einstein summation is implied.

The result in Eq.~\eqref{eq:Darcy1} is the fluid flux that arises solely from pressure changes, but it does not account for the fluid flow induced directly by the  thin filaments moving relative to the thick ones during contraction or relaxation. This movement has no effect on any of the lateral fluid fluxes $U_r$, nor on the axial flow in region 1,  $U_z^{(1)}$. In the thin region,  {the absence of stationary thick filaments means that the moving thin filaments pull the entirety of the fluid along with them, and hence }$U_z^{(3)}$ is decreased by a value  {equal to the porosity }$\Phi^{(3)}$. In the overlap region,  {the stationary thick filaments partially resist the flow induced by the moving thin filaments, and }$U_z^{(2)}$ decreases by a value $\beta\Phi^{(2)}$, where the traction parameter $\beta$ is a dimensionless constant between $0$ and $1$ that is determined entirely by the geometry of the cell, and calculated similarly to the axial permeabilities, in Appendix~\ref{appendix:beta}. Together with Eq.~\eqref{eq:Darcy1}, this gives us modified, final Darcy equations
\begin{equation}\label{eq:U}
U_z^{(j)} = -\gamma^{(j)}\Phi^{(j)} - k_{\|}^{(j)}\frac{\partial P^{(j)}}{\partial z}, \qquad U_r^{(j)} = -k_{\perp}^{(j)}\frac{\partial P^{(j)}}{\partial r},
\end{equation}
where $\gamma^{(1)} = 0, \gamma^{(2)} = \beta, \gamma^{(3)} = 1$ {capture the effects of the moving thin filaments in each region}.

\subsubsection{Pressure equation.}
{Since the fluid flow is incompressible, it follows that the same is true for the fluid flux and thus $\nabla\cdot\mathbf{U} = 0$.} Substituting the two Darcy equations from Eq.~\eqref{eq:U} into the incompressibility condition leads to  the governing equation for the Darcy pressure field $P$ as
\begin{equation}
\frac{1}{r}\frac{\partial}{\partial r}\left(r\frac{\partial P^{(j)}}{\partial r}\right) + \frac{k_{\|}^{(j)}}{k_{\perp}^{(j)}}\frac{\partial ^2P^{(j)}}{\partial z^2}= 0.
\label{eq:P}
\end{equation}
This is accompanied by relevant boundary conditions, as discussed below. Note that although the interstitial pressure $p$ is harmonic, as required by Stokes flow,  the Darcy pressure $P$ is not exactly harmonic; however, by a suitable rescaling of $z$, we would recover Laplace's equation and  so, with appropriate boundary conditions,  we deduce that $P$ has a unique solution.

\subsubsection{Boundary conditions for the full sarcomere model.}
We first identify appropriate boundary conditions for the full sarcomere model of Fig.~\ref{Figure 1}D, so that we can approximate them in the Darcy model, and later perform full numerical computations. We may exploit axial symmetry to consider only the right half-sarcomere $z \ge 0$. The surfaces of the individual filaments are subject to no-slip conditions, i.e.~the fluid velocity $\mathbf{u}$ is equal to the velocity of the filament. The Z disk and M disk (if present) are protein structures that are much more densely packed than the three filament regions, so it is legitimate to  approximate a no-slip condition on the disks as well. Alternatively, when the M disk is absent, we should apply a symmetry condition at $z = 0$: normal velocity is zero, $u_z = 0$, and normal gradient of tangential velocity is also zero, $\partial u_r /\partial z = \partial u_\theta /\partial z = 0$. We model the sarcomere as being immersed in free fluid, so in our simulations we also apply a zero-stress far-field condition away from the sarcomere.

\subsubsection{Boundary and interface conditions for the Darcy model.}

We now determine the appropriate boundary conditions, based on the above, for the  {Darcy model of the sarcomere}. Being a rescaled Laplace equation, Eq.~\eqref{eq:P}  requires a single boundary condition at each boundary. We set $P = 0$ at $r = 1$  {to simulate the stress-free far-field condition}. {In particular, this can be understood by noting that the dense internal structure of the sarcomere is expected to generate much larger pressure gradients inside the sarcomere compared to outside, differing by a factor of $(R/l)^2$, where $R$ ($\mathcal{O}$($1~\mu$m)) is the sarcomere radius and $l$ ($\mathcal{O}$($10~$nm)) is the pore scale, according to scalings of the Stokes equations. This assumption is confirmed by the full numerical simulations discussed above. We} apply no-penetration, $\partial P/\partial z = 0$, at the disks. This may produce a non-zero tangential slip velocity, in contrast to the full boundary conditions; however, we will see that the overall effect of this turns out to be very small on the scale of the full sarcomere. 
 Meanwhile, if the M disk is absent, the two symmetry conditions are $u_z = 0$ and $\partial u_r/\partial z = 0$,  both of which can be achieved by setting $\partial P/ \partial z = 0$. In other words, the Darcy flow is the same regardless of whether the M disk is present or not. Finally, there is an additional regularity condition $\partial P/\partial r = 0$ at $r = 0$ that arises from exploiting azimuthal symmetry {in the cylindrical geometry}.
 
Writing these explicitly, we therefore need to solve Eq.~\eqref{eq:P} subject to
\begin{align}
\frac{\partial P^{(1)}}{\partial r} = \frac{\partial P^{(2)}}{\partial r} = \frac{\partial P^{(3)}}{\partial r} = 0  \qquad &at ~ r = 0 \label{bc0} ,\\
P^{(1)} = P^{(2)} = P^{(3)} = 0  \qquad &at ~ r = 1 \label{bc1}, \\
\frac{\partial P^{(1)}}{\partial z} = 0 \qquad &at ~ z = 0 \label{bc2},\\
\frac{\partial P^{(3)}}{\partial z} = 0 \qquad &at ~ z = 1 \label{bc3}.
\end{align}

There remains the task of identifying the interface conditions between adjacent regions of the sarcomere. Assuming that the Darcy equations hold everywhere,  we identify the change in pressure across an interface by performing a force balance. We integrate the axial flux equation in a small region about the interface at $z = L_k$, giving
\begin{equation}
\left[P\right]_{L_k - \epsilon}^{L_k + \epsilon} = -\int_{L_k - \epsilon}^{L_k + \epsilon} \frac{U_z + \gamma\Phi}{k_{\|}}~dz.
\end{equation}
Taking $\epsilon \to 0$, we see that the change in pressure across the interface is zero;  {continuity of pressure is a standard condition for interfaces in porous media~\cite{Guyon_2015, Saffman_1971}}. Meanwhile, volume conservation gives continuity of axial flux at the interface between regions 2 and 3  {($z = L_2$)}. In contrast,  the interface between regions 1 and 2  {($z = L_1$)} is more subtle. As the sarcomere contracts, the interface between regions 1 and 2 moves leftwards. This moving boundary produces an effective flux source, and the resulting increase in axial fluid velocity when moving from region 1 to region 2 is precisely $(1 - \Phi^{(3)})$, which is the volume fraction occupied by the thin filaments. The interface conditions are therefore
\begin{align}
-k_{\|}^{(2)}\frac{\partial P^{(2)}}{\partial z} + k_{\|}^{(1)}\frac{\partial P^{(1)}}{\partial z} = \beta\Phi^{(2)} + \left(1-\Phi^{(3)}\right) \quad &at ~ z = L_1 \label{bc4},\\
-k_{\|}^{(3)}\frac{\partial P^{(3)}}{\partial z} + k_{\|}^{(2)}\frac{\partial P^{(2)}}{\partial z} = \Phi^{(3)} - \beta\Phi^{(2)} \quad &at ~ z = L_2 \label{bc5},\\
P^{(1)} - P^{(2)} = 0  \quad &at ~ z = L_1 \label{bc6},\\
P^{(2)} - P^{(3)} = 0 \quad &at ~ z = L_2 \label{bc7}.
\end{align}

\begin{table*}
\small
\begin{tabular*}{\textwidth}{@{\extracolsep{\fill}}llll}
\hline
Parameter & Symbol & Value (nm) & Reference \\
\hline
Thick filament radius & $R_{thick}$ & 7 &~\cite{Huxley_1953, Hayes_Huang_Zobel_1971, Malingen_Hood_Lauga_Hosoi_Daniel_2021} \\
Thin filament radius & $R_{thin}$ & 5 &~\cite{Holmes_Popp_Gebhard_Kabsch_1990, Malingen_Hood_Lauga_Hosoi_Daniel_2021} \\
D10 spacing & $D_{10}$ & 45 &~\cite{Malingen_Asencio_Cass_Ma_Irving_Daniel_2020, Shimomura_Iwamoto_Doan_Ishiwata_Sato_Suzuki_2016, Malingen_Hood_Lauga_Hosoi_Daniel_2021} \\
Thick filament spacing & $2D_{10}/\sqrt{3}$ & 52 & \\
Thick half-filament length & $L_{thick}$ & 800 &~\cite{Shimomura_Iwamoto_Doan_Ishiwata_Sato_Suzuki_2016, Malingen_Hood_Lauga_Hosoi_Daniel_2021}\\
Thin filament length & $L_{thin}$ & 1000 &~\cite{Shimomura_Iwamoto_Doan_Ishiwata_Sato_Suzuki_2016, Malingen_Hood_Lauga_Hosoi_Daniel_2021} \\
Sarcomere radius & $R$ & 500 &~\cite{González-Morales_Xiao_Schilling_Marescal_Liao_Schöck_2019} \\
\hline
\end{tabular*}
\caption{Physical parameters (length scales) used across all sarcomere models, with corresponding references \cite{Huxley_1953, Hayes_Huang_Zobel_1971, Malingen_Hood_Lauga_Hosoi_Daniel_2021, Holmes_Popp_Gebhard_Kabsch_1990, Malingen_Asencio_Cass_Ma_Irving_Daniel_2020, Shimomura_Iwamoto_Doan_Ishiwata_Sato_Suzuki_2016, González-Morales_Xiao_Schilling_Marescal_Liao_Schöck_2019}; the thick filament spacing is given by the D10 spacing  {(horizontal spacing between two adjacent colums of thick filaments in Fig.~\ref{Figure 2})} multiplied by $2/\sqrt{3}$. {All values are given to 1 or 2 significant figures}\label{table1}
}
\end{table*}

\subsubsection{Solution for pressure.}
The solution to Eq.~\eqref{eq:P} in region $j$ (with $j=1,2,3$) that satisfies the boundary conditions in Eqs.~\eqref{bc0}-\eqref{bc1}  is obtained classically as the series
\begin{equation}
P^{(j)}(r,z) = \sum_{n=1}^\infty \left[A_{n}^{(j)}e^{G_{n}^{(j)}z} + B_{n}^{(j)}e^{-G_{n}^{(j)}z}\right]J_0(\lambda_{n} r),
\label{eq:pressure}
\end{equation}
where $J_0$ is the zero order Bessel function of the first kind, $\lambda_n$ is the $n$th zero of $J_0$, and $G_{n}^{(j)} = \lambda_n \sqrt{k_{\perp}^{(j)}/k_{\|}^{(j)}}$.  {This solution satisfies the boundary conditions Eqs.~\eqref{bc0}-\eqref{bc1} analytically;  the two remaining boundary conditions Eqs.~\eqref{bc2}-\eqref{bc3}, as well as the four interface conditions Eqs.~\eqref{bc4}-\eqref{bc7}, must be satisfied by appropriate choice of coefficients $A_n^{(j)}$ and $B_n^{(j)}$. Since there are three regions, this amounts to $6$ coefficients to be determined for each value of $n$.} By using the orthogonality condition between Bessel functions, we may apply the boundary and interface conditions Eqs.~\eqref{bc2}-\eqref{bc7} to obtain an appropriate $6 \times 6$ matrix equation for the six coefficients above, for each value of $n$ independently (see Appendix~\ref{appendix:boundary_and_interface_conditions}). Once we calculate the permeabilities and the traction parameter $\beta$, we easily solve these matrix equations numerically, and hence obtain all the coefficients. Of course, in order to compute the solution, Eq.~\eqref{eq:pressure}, we must truncate the series to some finite number of terms $N$,
\begin{equation}
P^{(j)}(r,z) \approx \sum_{n=1}^N \left[A_{n}^{(j)}e^{G_{n}^{(j)}z} + B_{n}^{(j)}e^{-G_{n}^{(j)}z}\right]J_0(\lambda_{n} r),
\label{eq:pressure_truncated}
\end{equation}
where, to reiterate, the permeabilities are incorporated via $G_n^{(j)} = \lambda_n\sqrt{{k_\perp^{(j)}}/{k_\|^{(j)}}}$, with $\lambda_n$ being the $n$th zero of the Bessel function $J_0$. In all results below, we set $N = 50$  {(see Appendix~\ref{appendix:boundary_and_interface_conditions} for justification)}.

 \subsection{Biophysical parameter values in biological sarcomeres}
\label{sec:actualvalues}

In order to validate the Darcy model, we must compare it with precise numerical computations. To this end, we identify here suitable values of the various biophysical parameters. We note that there is a wide range of biological parameters, observed in a variety of sarcomeres; below we will test the Darcy model in a minimal suite of configurations, for which we expect the worst agreement, and will next discuss the extension to more ideal configurations. 

\subsubsection{Fixed value parameters: Filament radii and spacing.}

{A review of the existing literature, examining both intact sarcomeres and isolated actin and myosin filaments, reveals that certain biophysical parameters are approximately constant across all sarcomeres \cite{Huxley_1953, Hayes_Huang_Zobel_1971, Malingen_Hood_Lauga_Hosoi_Daniel_2021, Holmes_Popp_Gebhard_Kabsch_1990, Malingen_Asencio_Cass_Ma_Irving_Daniel_2020, Shimomura_Iwamoto_Doan_Ishiwata_Sato_Suzuki_2016}.} The radii of both thick and thin filaments, as well as the D10 filament spacing,  {$D_{10}$, }which is defined as the horizontal distance between two adjacent columns of thick filaments in Fig.~\ref{Figure 2}, will be assumed to be constant. Consequently, the distance between any two adjacent thick filaments, given by ${2D_{10}}/{\sqrt{3}}$, is also constant. These values are listed in Table~\ref{table1}.

Geometrically, the thick and thin filament radii are dictated by the fixed size of the individual myosin and actin monomers, {respectively}, whilst the filament spacing is dictated by the distance over which myosin molecules can reach during the cross-bridge cycle; we can therefore also consider these parameters to be constant. Structural variation between sarcomeres, such as different packing ratios, can affect these values to an extent; we selected values, consistent with the larger body of literature, to match those used in previous numerical work~\cite{Malingen_Hood_Lauga_Hosoi_Daniel_2021}. 

In what follows, we will also set the  {dimensional} muscle contraction speed $V$  to $1000~$nm/s, which is the observed order of magnitude for sarcomere contraction~\cite{ter_Keurs_Diao_Deis_2010, Rodriguez_Hunter_Royce_Leppo_Douglas_Weisman_1992, Shankar_Mahadevan_2024a}. Obviously, due to the linearity of the (viscous) physical system, all flow will instantaneously scale linearly with the chosen value of $V$. 

\subsubsection{Extreme value parameters: Filament lengths and sarcomere length; sarcomere radius.}

The remaining parameters could be selected from a wide range of biologically observed values. Filament lengths have been reported to vary through a factor of at least around $5$, though there is a systematic linear relationship between thick and thin filament lengths~\cite{Shimomura_Iwamoto_Doan_Ishiwata_Sato_Suzuki_2016}. 
 Sarcomere radius can also vary significantly~\cite{González-Morales_Xiao_Schilling_Marescal_Liao_Schöck_2019}.  The Darcy model is expected to be most valid when there are large relative length scales of variation, and in order to validate it we will  focus on the worst-case scenario, i.e.~situations of shortest observed length scales.  We thus set sarcomere radius and filament lengths (and consequently overall sarcomere length) to take their smallest observed values; these   have been added to Table~\ref{table1}.

\subsubsection{Variable parameters: Contractions; M disk; packing ratio.}

With all biophysical parameters set, we  vary in our investigation the level of contraction, $\alpha \equiv 1 - L_1/L_2$, between states of extreme contraction ($\alpha$ large) and extreme relaxation ($\alpha$ small), thereby considering the full range of possible values for the lengths of the three regions. We expect states of extreme contraction or relaxation, both of which lead to the presence of  short regions, to produce the least accurate agreement  {with the full numerical computations}. In reality, sarcomeres do not typically contract over the full range, although contractions of hundreds of nanometres are observed even in   small sarcomeres~\cite{ter_Keurs_Diao_Deis_2010, Rodriguez_Hunter_Royce_Leppo_Douglas_Weisman_1992}. We also consider {the} presence or absence of the M disk, and cases of differing packing ratios.

 {\subsubsection{Permeabilities and traction parameter $\beta$.}
Thanks to the aforementioned constant filament radii and spacing, the traction parameter $\beta$ and the six  {(dimensional) }Darcy permeabilities are constant for each packing ratio. While $\beta$ is a truly dimensionless parameter, the dimensional permeabilities each have dimensions of length squared, and must be appropriately non-dimensionalised to be used in Eq.~\eqref{eq:pressure_truncated}. Specifically, each dimensional $k_\|^{(j)}$ (being a property of axial flow) should be divided by $L^2$, and each dimensional $k_\perp^{(j)}$ (being a property of lateral flow) by $R^2$, for dimensional sarcomere length $L$ and radius $R$. Therefore, despite each dimensional permeability being independent of $L$ and $R$, their non-dimensional counterparts in Eq.~\eqref{eq:pressure_truncated} are not. As such, in order to present the permeabilities in a succinct, dimensionless form, that does not depend on $L$ or $R$, we introduce new non-dimensionalisations specifically for the triangular cells of Fig.~\ref{Figure 2}F-H within which the permeabilities are calculated. All lengths are scaled with the triangle height, $D_{10}/\sqrt{3}$, so that the triangles now have a dimensionless height of $1$. The resultant dimensionless permeabilities have been computed in  {Appendix~\ref{appendix:axial_permeabilities} and \ref{appendix:lateral_permeabilities}, with $\beta$ being calculated in Appendix~\ref{appendix:beta},} and these are listed in Table~\ref{table2} for packing ratios of 2 (left column), 3 (middle) and 5 (right). To restore the permeabilities to their dimensional values, they must be multiplied by $D_{10}^2/3$. In order to use the permeabilities in Eq.~\eqref{eq:pressure_truncated}, each $k_\|^{(j)}$ must then be divided by $L^2$ and each $k_\perp^{(j)}$ must be divided by $R^2$. Therefore overall, to use the permeabilities in Table~\ref{table2} within Eq.~\eqref{eq:pressure_truncated}, each $k_\|^{(j)}$ must be multiplied by ${D_{10}^2}/{3L^2}$, and each $k_\perp^{(j)}$ must be multiplied by ${D_{10}^2}/{3R^2}$. In contrast, $\beta$ is a truly dimensionless quantity that requires no such treatment.}

\begin{table}[t]
\small
\begin{tabular*}{0.48\textwidth}{@{\extracolsep{\fill}}llll}
\hline
\multirow{2}{*}{Variable} ~~~~~~ & \multicolumn{3}{c}{Packing ratio} \\ \cmidrule{2-4} & ~~~~~ 2 ~~~~~  & ~~~~~ 3 ~~~~~ & ~~~~~ 5 ~~~~~ \\
\hline
$k_\|^{(1)}$ & 0.3729 & 0.3729 & 0.3729 \\
$k_\|^{(2)}$ & 0.0728 & 0.0433 & 0.0247 \\
$k_\|^{(3)}$ & 0.2038 & 0.1071 & 0.0759 \\
$\beta$ & 0.6175 & 0.7127 & 0.7238 \\
$k_\perp^{(1)}$ & 0.1864 & 0.1864 & 0.1864 \\
$k_\perp^{(2)}$ & 0.0362 & 0.0212 & 0.00618 \\
$k_\perp^{(3)}$ & 0.0938 & 0.0435 & 0.00737 \\
\hline
\end{tabular*}
\caption{Computed  {dimensionless} axial and lateral permeabilities in the three regions{, as well as the traction parameter $\beta$,}  for packing ratios of 2, 3 and 5, using a triangle with height $1$, see Fig.~\ref{Figure 2} {and Appendix~\ref{appendix:axial_permeabilities}, \ref{appendix:beta}, \ref{appendix:lateral_permeabilities}}.  {To use these values in Eq.~\eqref{eq:pressure_truncated}, each stated value of $k_\|^{(j)}$ must be multiplied by ${D_{10}^2}/{3L^2}$, and each $k_\perp^{(j)}$ must be multiplied by ${D_{10}^2}/{3R^2}$, where $L$ and $R$ are the dimensional sarcomere length and radius, respectively (see main text and Appendix~\ref{appendix:axial_permeabilities}, \ref{appendix:lateral_permeabilities})}. All geometrical parameters are taken from Table~\ref{table1}\label{table2}
}
\end{table}

\section{Fluid flux in sarcomeres}
\label{sec:results}

In this section, we apply our modelling approach to compute the fluid flux in and out of contracting sarcomeres. Specifically, we compute the predictions of our Darcy model and compare them with results of full numerical computations obtained using the finite element method (FEM) within COMSOL Multiphysics~\cite{COMSOL}, wherein the entire geometry of the sarcomere is fully specified and the flow past the array of filaments is explicitly calculated (see Appendix~\ref{appendix:FEM_computations} for details).

It should be emphasised that the two approaches (Darcy vs FEM)  are fundamentally different, with the latter considering conventional fluid properties, and the former considering only the bulk fluid properties; as such, we will need to average the FEM data in order to compare with the Darcy model. Note that the comparisons will be made using dimensional variables, to help with application to biological systems. As we will see, the  Darcy model is able to accurately reproduce the bulk fluid flow of the precise FEM computations in an  averaged sense, 
serving as a highly accurate, and  far more time-efficient and less resource-intensive, method of calculating the flow within the sarcomere.

\subsection{Illustrative example}

\subsubsection{Fluid flow.}

\begin{figure}[t]
\centering
\includegraphics[width = 0.48\textwidth]{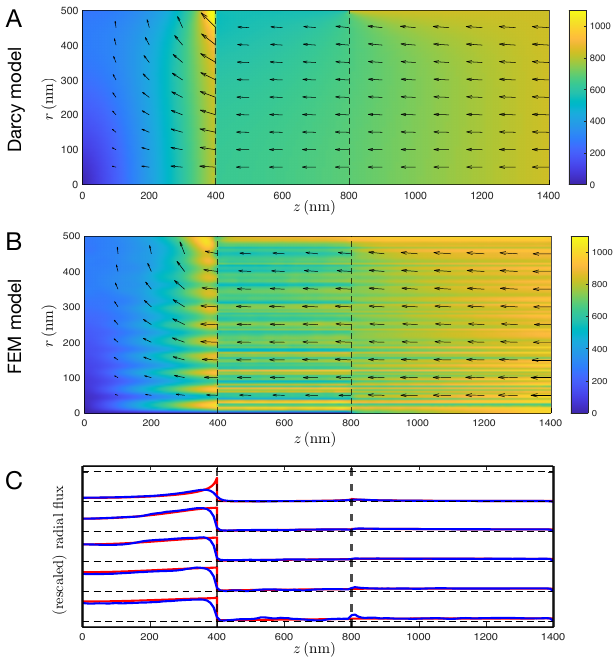}
\caption{Comparisons of fluid flow between the Darcy and FEM models, for a sarcomere at $50\%$ contraction with a packing ratio of 5 in the absence of an M disk. (\textbf{A}, \textbf{B}) Heat map of fluid flux with arrows indicating direction for Darcy (A) and FEM (B) models.  {Note that in A, the maximum flux magnitude (found at the radial boundary, at the interface between regions 1 and 2) is approximately $1600~$nm/s, whereas the colour scale has a maximum value of $1100~$nm/s; this is to improve readability of A and B, and less than $0.1$\% of the domain surpasses $1100~$nm/s.} (\textbf{C}) Plot of radial flux against $z$ for the Darcy (red) and FEM (blue) models, with $r = 100, 200, 300, 400, 500~$nm from bottom to top. Radial fluxes have been individually rescaled to have the same maximum value of radial Darcy flux, to improve readability.}
\label{Figure 3}
\end{figure}

We start our comparisons between the two models by focusing on the particular case of a sarcomere with a packing ratio of 5 in the absence of an M disk  {(typical of invertebrates~\cite{Page_1965, Peachey_Huxley_1962, Craig_Padron_2004, Shimomura_Iwamoto_Doan_Ishiwata_Sato_Suzuki_2016})}, at $50\%$ contraction, i.e. $\alpha = 1 - L_1/L_2 = 0.5$. To carry out a detailed comparison between the Darcy and FEM approaches, we azimuthally average the FEM data and plot the flux magnitude as a heat map, with arrows indicating direction of flux. The results are shown in Fig.~\ref{Figure 3} with Darcy predictions in Fig.~\ref{Figure 3}A, and FEM numerics in Fig.~\ref{Figure 3}B. 

We observe excellent agreement between the two models, in terms of the magnitude, direction and spatial distribution of the flow. The striations in the FEM model, most visible in the overlap region, correspond to the presence and absence of filaments; these filaments cause fluctuation in the axial fluid flux (but not the radial fluid flux) which are effectively averaged out by the Darcy model.

\subsubsection{Radial flux.}
Since the  heat maps in Fig.~\ref{Figure 3}A-B alone do not offer a  proper quantitative comparison between the two models, we next compute the rescaled radial flux against $z$ at a number of fixed values of $r$ and compare with the average of the computational results   in Fig.~\ref{Figure 3}C.  The Darcy model accurately captures the radial fluxes across nearly all values of $r$.

Some small quantitative differences can be observed. For very small $r$, the model deteriorates somewhat; this represents only a few percent of the entire sarcomere volume and is not of practical  concern. Some discrepancies are also seen for very large $r$, as the periodic pattern begins to terminate. Specifically, whilst the Darcy model considers a porous medium in a perfect cylindrical geometry, the FEM model explicitly considers a hexagonal lattice that must be fitted inside said cylinder. As such, the lattice must be terminated as it approaches the outer radial boundary. As a result, within the FEM model, there are small regions near the outer radial boundary but still within the cylinder for which the hexagonal lattice of filaments is absent.

Considering variation of $z$, the most noticeable discrepancies between the porous media approach and the full simulations appear near the interfaces between the regions of the sarcomere, where the assumptions necessary for Darcy flow break down, and near the no-slip condition at $z = L$, which cannot be precisely satisfied by the Darcy model. These create short regions of disagreement, across thicknesses that are about the size of the hexagonal cells. 

\begin{figure}[t]
\centering
\includegraphics[width = 0.48\textwidth]{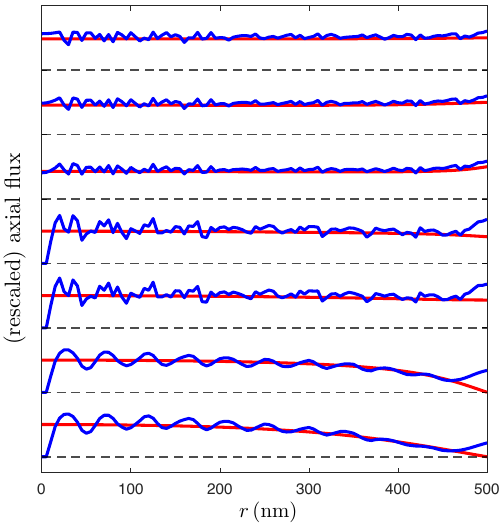}
\caption{Comparisons of fluid flow between the Darcy and FEM models, for a sarcomere at $50\%$ contraction with a packing ratio of 5 in the absence of an M disk. Plot of axial flux against $r$ for the Darcy (red) and FEM (blue) models, with $z = 100$, $300$, $500$, $700$, $900$, $1100$, $1300~$nm from bottom to top. Axial fluxes have been individually rescaled to have the same maximum value of axial Darcy flux, to improve readability.}
\label{Figure 4}
\end{figure}

\subsubsection{Axial flux.}
We next consider the  {axial} fluxes through the sarcomere. Given the very good agreement  {of radial flux illustrated in Fig.~\ref{Figure 3}C}, it follows from mass conservation that the axial fluid fluxes should also  agree in an averaged sense. This is confirmed directly in Fig.~\ref{Figure 4},  {where the azimuthally averaged FEM axial flux  {and the Darcy axial flux are plotted as functions} of $r$ for select values of $z$. Overall we also find excellent agreement at almost all values of $r$ and $z$, with notable exceptions as discussed above. In particular, note that the fluctuations of the FEM axial flux caused by the presence of the filaments are averaged out by the Darcy model}. 

\subsubsection{Lagrangian transport.}

\begin{figure}[t]
\centering
\includegraphics[width= 0.4\textwidth]{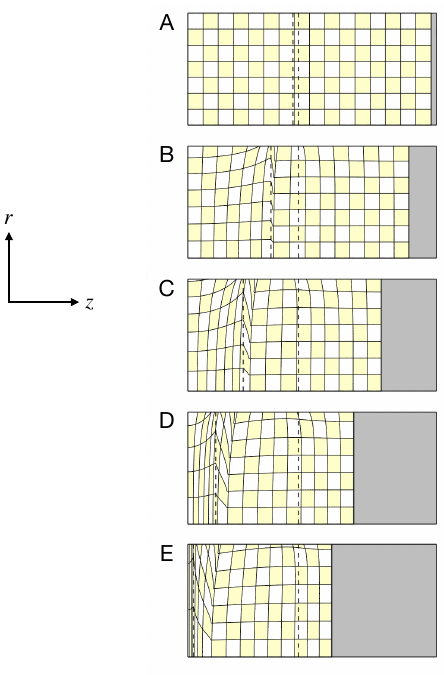}
\caption{Lagrangian deformation of marked fluid particles throughout a sarcomere contraction, shown for contractions of $\alpha = 5 \%$ (A), 25\% (B), 50\% (C), 75\% (D) and $95 \%$ (E). Dashed {black} lines indicate interfaces between the three regions.}
\label{Figure 5}
\end{figure}

Given this excellent agreement, we can now use the Darcy model to compute and consider the Lagrangian deformation of fluid particles within the sarcomere. We numerically integrate (via a first-order forward Euler scheme) the  {interstitial fluid velocity, $\mathbf{u} = \mathbf{U}/\Phi$,} of the Darcy model to  determine  the deformation of marked fluid elements over the course of a contraction. The results are illustrated in Fig.~\ref{Figure 5} for sarcomere contractions of $\alpha = 5 \%$ (A), 25\% (B), 50\% (C), 75\% (D) and $95 \%$ (E). 

 {We  observe that fluids in the thin and overlap regions typically undergo very little deformation. Meanwhile the fluid in the thick region undergoes large displacements and deformations in both the axial and radial directions, with significant axial compression and radial stretching of the fluid parcels, and large radial efflux out of the sarcomere. Due to the reversibility of Stokes flow, this fluid will be drawn back (deterministically) into the sarcomere as it relaxes. {This may help to draw in useful substrates (such as ATP) during relaxation~\cite{Cass_Williams_Irving_Lauga_Malingen_Daniel_Sponberg_2021}. Furthermore, various metabolic processes occur within the sarcoplasm, involving a variety of substrates with uses and effects both inside and outside the sarcomere~\cite{Hargreaves_Spriet_2020}. Fluid flow may beneficially redistribute these substrates, and in particular may help rid the sarcomere of waste products (such as lactate resulting from anaerobic respiration, and excess ADP resulting from hydrolysis of ATP over prolonged activity) during contraction.} Interestingly, the axial compression of the fluid is insufficient to keep the fluid parcels, that begin in the thick region, in the thick region; the growing overlap region absorbs much of this fluid. Similarly, much of the fluid that begins in the thin region is pushed into the growing overlap region. Both of these are likely to help advect useful substrates, such as ATP and calcium ions needed for cross-bridging~\cite{Zot_Potter_1987, Solaro_Rarick_1998, Gordon_Homsher_Regnier_2000}, into the overlap region. Importantly, advection of ATP is likely to be highly impactful to sarcomere function, since the internal structure of the sarcomere has been shown to impede ATP diffusion by several orders of magnitude compared to free cytosol~\cite{Selivanov_Krause_Roca_Cascante_2007a, Alekseev_Guzun_Reyes_Pison_Schlattner_Selivanov_Cascante_2016}.}

\begin{figure}[t]
\centering
\includegraphics[height = 0.6\textwidth]{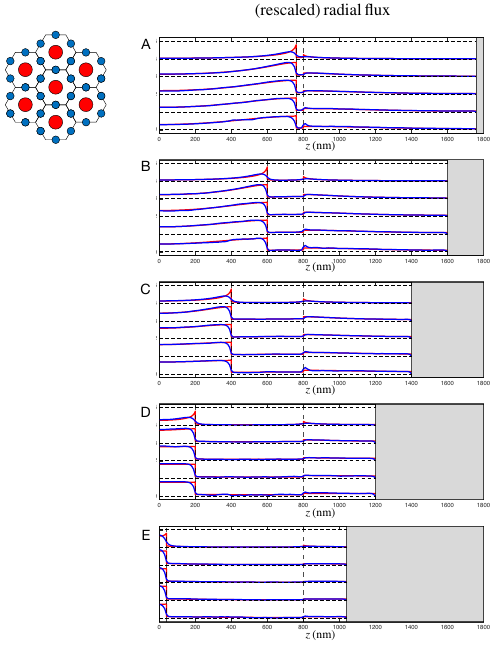}
\caption{Radial fluid flux in a sarcomere with a packing ratio of 3 (cross-section of overlap region illustrated on top left) in the absence of an M disk at various stages of contraction, $\alpha = 5\%$ (A), 25\% (B), 50\% (C), 75\% (D), 95\% (E), showing the Darcy model (red) against the FEM model (blue). Each figure shows five pairs of data sets, at $r = 100,~200,~300,~400,~500~$nm from bottom to top.}
\label{Figure 6}
\end{figure}

\subsection{Analysis and comparisons of a variety of sarcomeres}

Having established agreement between the Darcy model and full numerical simulations  {in the above illustrative example}, we investigate in this subsection the impact of varying the nature of the sarcomere. Specifically, we show a similar  comparison in the case of a sarcomere with a packing ratio of 3 in the absence of an M disk at various stages of contraction  in Fig.~\ref{Figure 6}; this is the case characteristic of invertebrate organisms \cite{Page_1965, Peachey_Huxley_1962, Craig_Padron_2004, Shimomura_Iwamoto_Doan_Ishiwata_Sato_Suzuki_2016}. We also show a similar result for a sarcomere with a packing ratio of 2 in the presence of an M disk in Fig.~\ref{Figure 7}, here a situation relevant to vertebrate sarcomeres \cite{Knappeis_Carlsen_1968, Luther_Squire_1978, Luther_Crowther_1984, Crowther_Luther_1984, Craig_Padron_2004, Stenger_Spiro_1961, HOYLE_1967, Shimomura_Iwamoto_Doan_Ishiwata_Sato_Suzuki_2016}.

\begin{figure}[t]
\centering
\includegraphics[height = 0.6\textwidth]{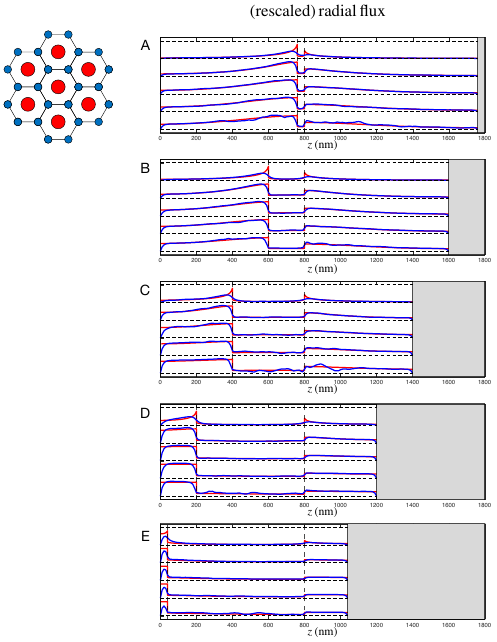}
\caption{Radial fluid flux in a sarcomere with a packing ratio of 2 (cross-section of overlap region illustrated on top left) in the presence of an M disk at various stages of contraction, $\alpha = 5\%,~25\%,~50\%,~75\%,~95\%$ from top to bottom, showing the Darcy model (red) against the FEM model (blue). Each figure shows five pairs of data sets, at $r = 100,~200,~300,~400,~500~$nm from bottom to top.}
\label{Figure 7}
\end{figure}

   In both cases, we again observe   overall excellent agreement between the two models. When the M disk is present (Fig.~\ref{Figure 7}), some discrepancies are seen in its vicinity, all the more present if the thick region is very small (limit of large contractions). Despite this, the agreement between the Darcy approach and the averaged numerics is essentially perfect at most values of $r$ and $z$.

 {We are now able to compare the properties of the flow between the different sarcomeres.} By examining Figs.~\ref{Figure 3}, \ref{Figure 6} and \ref{Figure 7}, we see that the radial efflux is low within the overlap region, owing to the low permeability caused by the filaments,  {but is typically greater in the thick and thin regions, a consequence of the higher permeabilities.} Generally speaking, the thick region contributes the majority of radial flow out of the sarcomere - the sarcomere contracts, and the thin filaments invade the thick region,  thus pushing fluid out of the sarcomere. A similar argument of thick filaments invading the thin region (after an appropriate change of frame) explains the significant radial outflow occurring in the thin region. The radial outflow in the thick region is most significant for higher packing ratios; indeed, as the packing ratio increases, the relative importance of the thin filaments increases. As such, the `plunger' effect of the thin filaments invading the thick region is effectively increased, leading to a more significant radial efflux in the thick region. Simultaneously, this increase in packing ratio reduces the permeability in the thin region, thereby reducing the radial efflux there. Predictably, this {`plunger'} effect is most evident near the interfaces with the overlap region.

 Overall, we see that the Darcy model  accurately predicts the radial flux, and hence all fluid properties, for all levels of contractions and sarcomere geometries. This agreement also allows us to confirm,  {\textit{a posteriori}}, that the methods by which the permeabilities were calculated (values in Table~\ref{table2}) were correct, and hence, again in an averaged sense, the Darcy model accurately determines the pressure and the interstitial flow profiles between the filaments.

\section{Discussion}\label{sec:discussion}
\subsection{Summary}
The fluid flow past the many interdigitating filaments of the sarcomere requires significant computational resources to calculate accurately. In this paper, we demonstrated how to  derive a Darcy model that considers the sarcomere as an anisotropic porous medium, and we compare its predictions with those of full numerical (FEM) simulations. 

The model gives a simple governing equation for the bulk pressure $P$ (Eq.~\eqref{eq:P}) with further equations to obtain the bulk fluid flux from this pressure (Eq.~\eqref{eq:U}). The boundary conditions valid in a full computational (FEM) model can be approximated in the Darcy model, and force and flux balances allow for appropriate interface conditions between the regions of the sarcomere. The final solution for $P$ is a truncated series (Eq.~\eqref{eq:pressure_truncated}) whose coefficients are easily calculated from these boundary and interface conditions, by exploiting the orthogonality condition between Bessel functions. In order to do this, we calculate six permeabilities and the traction parameter $\beta$ (Table~\ref{table2}), which is done by adapting classical semi-analytic schemes to triangles representing $1/12$ of a hexagonal cell. The calculated values for the permeabilities and $\beta$ are then fed into the equations for the coefficients (Appendix~\ref{appendix:boundary_and_interface_conditions}). Having determined these coefficients, we can evaluate $P$ and $\mathbf{U}$ at any given point in the sarcomere practically instantly. 

Overall, we obtain excellent agreement between the Darcy and FEM models, with minor exceptions near the radial centre and exterior radial boundary of the sarcomere, as well as near interfaces and disks. The Darcy model therefore allows us to calculate, with surprising accuracy, the fluid flow within the sarcomere, in a vastly reduced amount of time. Indeed, the FEM model required a multi-processor machine with specialised software, taking around 24 hours to compute a particular solution. Meanwhile{, once the permeabilities are calculated}, the Darcy model can be computed using basic software in {less than a second} on a simple laptop computer; {the permeabilities themselves can be calculated in less than a minute on the same device.}

{\subsection{Potential Applications}}}
In particular, this allows us to calculate the fluid flow easily in a continuous range of contractions, enabling us to accurately determine the time-evolution of the fluid within the sarcomere. This allows us to study, both quantitatively and qualitatively, the behaviour of the fluid flow, and consider how this may affect the function of the sarcomere, such as by the transport of substrates beyond that achieved by diffusion alone. {A demonstration of this phenomenon is illustrated in Appendix~\ref{appendix:ARD}, where preliminary results on substrate advection, reaction and diffusion are shown.

{Another potential application is in the calculation of viscous drag or viscous energy dissipation as a result of fluid flow. Importantly, these must be calculated at the pore scale, analysing the semi-analytic flow profile to obtain a linear (though anisotropic) relationship between pressure gradient (or fluid flux) and viscous drag, and a quadratic relationship between pressure gradient (or fluid flux) and viscous energy dissipation. These can then be integrated over the sarcomere to obtain an expression for the total drag force and total rate of doing work. Alternatively, order-of-magnitude estimates can be obtained using scaling arguments, with viscous drag per unit filament surface area on the order of $\mu u/l$, and viscous dissipation per unit fluid volume on the order of $\mu u^2/l^2$. Here $u$ is a characteristic fluid velocity ($10^{-6}~$m/s~\cite{ter_Keurs_Diao_Deis_2010, Rodriguez_Hunter_Royce_Leppo_Douglas_Weisman_1992, Shankar_Mahadevan_2024a}) and $l$ is the relevant length scale (i.e. the pore scale, $10^{-8}~$m). Even using an upper estimate for sarcoplasm viscosity (being a few tens of times that of water, $\mathcal{O}(10^{-2})~$Pa~s \cite{Kalwarczyk_et_al_2011, Kwapiszewska_et_al_2020}) we find that the viscous drag experienced by each filament is at most $\mathcal{O}(10^{-14})~$N. This figure was also classically obtained by Huxley, who additionally noted that the total viscous drag experienced by the filaments in the sarcomeres is several orders of magnitude smaller than the forces muscles generate, and so is unlikely to be of consequence~\cite{Huxley_1980}. Viscous energy dissipation per unit volume can be estimated either using the aforementioned direct scaling argument, or simply by multiplying the viscous drag force experienced by each filament by the contraction speed, and accounting for the number of filaments per unit volume, and is found to be at most $\mathcal{O}(10^2)~$W/m$^3$. This is several orders of magnitude smaller than the rate of energy metabolism from ATP hydrolysis, being around $10^5~$W/m$^3$~\cite{Hargreaves_Spriet_2020, Wackerhage_Hoffmann_Essfeld_Leyk_Mueller_Zange_1998}, and so is also unlikely to be of consequence.}}

\subsection{Versatility of semi-analytic solutions for flow within periodic cells}
The semi-analytic methods of calculating the fluid flow, and hence the fluid flux and permeability, through the periodic cells are highly adaptable. For example, whilst only three possible packing ratios are shown in Fig.~\ref{Figure 2}, packing ratios of 4, 6 and even 12 have been reported in biological sarcomeres~\cite{Shimomura_Iwamoto_Doan_Ishiwata_Sato_Suzuki_2016}. It is a simple matter to adapt the above methods to these more exotic packing ratios, simply by placing thin filaments appropriately along the lower edge of the triangle (Fig.~\ref{Figure 2}F-H). Furthermore, the general method can be adapted to other arrangements of filaments, such as a square arrangement (e.g. as considered by Sangani \& Acrivos~\cite{Sangani_Acrivos_1982}) by splitting the traditional periodic cell into appropriate right-angled triangles. In the case of a square arrangement, these would be isosceles right-angled triangles. Since only one side of the triangle is exposed to the boundary of the larger periodic cell (the other two are simply lines of symmetry within the cell) only that boundary can contain auxiliary filaments, making it a theoretically  easy (though sometimes numerically cumbersome) matter to alter the numerical boundary conditions as appropriate in the presence or absence of auxiliary filaments. Lastly, there is nothing in our analyses that demands the thin filaments are of equal size, or that all of the thin filaments are moving when calculating the traction parameter $\beta$. This is not relevant for sarcomeres, but could prove useful in other situations requiring calculation of the flow past a complex periodic array of (active) cylinders. 

\subsection{Limitations of the model sarcomere}

The model studied in this paper (Fig.~\ref{Figure 1}D)  is a mathematical and physical  simplification of reality that allows us to demonstrate the effectiveness of the Darcy approach to muscle flows, and a number of potential improvements could be proposed. 

Whilst myosin and actin are the two most abundant proteins of the sarcomere, composing approximately 50\% and 20\% of myofibrillar {protein} mass respectively~\cite{Huxley_Hanson_1957_1, Hanson_Huxley_1957_2}, we did not include the giant protein titin found in vertebrates, which makes up around 10\% of the myofibrillar {protein} mass~\cite{Craig_Padron_2004, Gregorio_Granzier_Sorimachi_Labeit_1999}. Titin has a diameter of around 4 nm, and each titin molecule runs from the M disk to the Z disk, via the thick filaments~\cite{Tskhovrebova_Trinick_2003, Higuchi_Nakauchi_Maruyama_Fujime_1993a, Trinick_Knight_Whiting_1984, Wang_Ramirez-Mitchell_Palter_1984}. Titin can also bind to actin in certain  {places}, though some of these are uncertain~\cite{Nagy_Cacciafesta_Grama_Kengyel_Málnási-Csizmadia_Kellermayer_2004, Yamasaki_Berri_Wu_Trombitás_McNabb_Kellermayer_Witt_Labeit_Labeit_Greaser_et_al_2001, Bianco_Nagy_Kengyel_Szatmári_Mártonfalvi_Huber_Kellermayer_2007, Kulke_Fujita-Becker_Rostkova_Neagoe_Labeit_Manstein_Gautel_Linke_2001, Linke_Ivemeyer_Labeit_Hinssen_Rüegg_Gautel_1997, Linke_Kulke_Li_Fujita-Becker_Neagoe_Manstein_Gautel_Fernandez_2002, Trombitas_Granzier_1997, Dutta_Tsiros_Sundar_Athar_Moore_Nelson_Gage_Nishikawa_2018}, and it is unclear what precise shape the titin molecules take in the thin region. Adding this to the fact that there are  {believed to be six} titin molecules per myosin half-filament~\cite{Liversage_Holmes_Knight_Tskhovrebova_Trinick_2001}, we see that the presence of titin may have a non-trivial effect on fluid flow within the thin region, and this could be the basis of future work.

A further complication in the specific case of vertebrate sarcomeres is that the thin region is often not a neat continuation of the hexagonal geometry in the overlap region. The thin filaments can become disorganised within the thin region, and appear to ultimately bind with the Z disk in a square arrangement~\cite{Craig_Padron_2004, Tskhovrebova_1991, Squire_Al‐khayat_Knupp_Luther_2005}. {The resultant uncertainty in the permeabilities within the thin region motivates a demonstration of how the model can be applied to a continuous range of permeabilities, which could be measured experimentally, without the need to resolve the precise physical structure within the thin region; we refer the reader to  Appendix~\ref{appendix:thin_region} for further details.}

 {We finally note that in this work we do not focus on cardiomyocytes (i.e.~heart muscle cells) that  contain branching myofibrils whose radii can be as small as 
$200~$nm~\cite{Burbaum_Schneider_Scholze_Böttcher_Baumeister_Schwille_Plitzko_Jasnin_2021, Moore_Ruska_1957, Linke_Popov_Pollack_1994, Oda_Yanagisawa_2020}, significantly smaller than the $500~$nm considered in Table \ref{table1}. This smaller length scale could impact the validity of the Darcy model, and accounting for this should be the focus of future work.}   

We hope that the simplicity of  our Darcy approach, and its agreement with full simulations, will encourage the further development of porous media models for muscle hydrodynamics.
 
\section*{Conflicts of interest}
There are no conflicts of interest to declare.

\section*{Code and data availability}
 The MATLAB code and processed COMSOL data needed to compute and verify the Darcy flow, and produce Figs.~\ref{Figure 3} - \ref{Figure 7}, have been made freely available on GitHub \cite{GitHub}.

\section*{Acknowledgements}
This project has received funding from the European Research Council (ERC) under the European Union's Horizon 2020 research and innovation programme  (grant agreement 682754 to EL), and from the European Commission's Erasmus+ programme. We  thank Sage Malingen, Richard Tyser, Grae Worster {and Marco Vona} for useful discussions and advice.

\appendix
\section{Calculating axial permeabilities}
\label{appendix:axial_permeabilities}

\begin{figure}[t]
\centering
\includegraphics[height = 4cm]{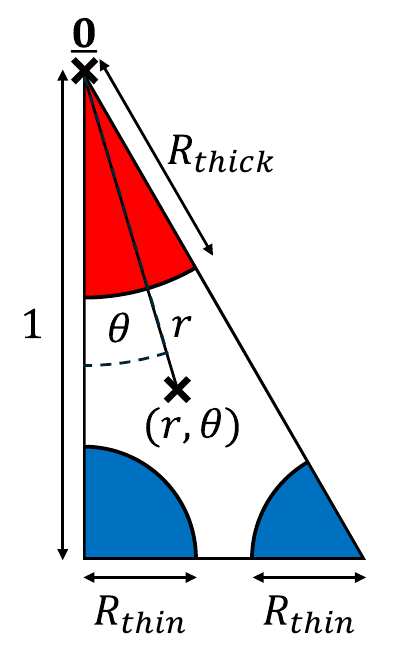}
\caption{Triangular domain, $1/12$ of a periodic hexagonal cell, on which the axial and lateral flow profiles are calculated, shown here for the overlap region with a packing ratio of 5 (see Fig.~\ref{Figure 2}). The triangle's original height of $D_{10}/\sqrt{3}$ is rescaled (using a different non-dimensionalisation to that used for the Darcy model) to $1$. Polar coordinates $(r, \theta)$, distinct from those used for the Darcy model, are shown, with rescaled thick and thin filament radii $R_{thick}$ and $R_{thin}$ indicated.}
\label{Figure 8}
\end{figure}

In this first Appendix, we apply and adapt a semi-analytic method to calculate the axial flow induced by a constant axial pressure gradient~\cite{Sparrow_Loeffler_1959}. The cross-section of the sarcomere seen in Fig.~\ref{Figure 2}A, of hypothetical radius $200~$nm,  consists of a hexagonal filament lattice, as displayed in Fig.~\ref{Figure 2}C-E, which shows packing ratios of thin actin filaments to thick myosin half-filaments of 2, 3 and 5 respectively; note that  this  illustrates the lattice within the overlap region, where both thick and thin filaments are present. The lattices in the thick and thin regions are obtained simply by removing the appropriate filaments. Therefore, each region exhibits a hexagonal lattice. These hexagonal cells can themselves be split into 12 right-angled triangles, seen in Fig.~\ref{Figure 2}F-H respectively{, and reproduced for a packing ratio of 5 in Fig.~\ref{Figure 8}}. Note that, for all the packing ratios presented, and any region of the sarcomere, the filaments present as sectors occupying the vertices of these triangles.

These triangles within the lattice have dimensional height $H {= D_{10}/\sqrt{3}}$, but to ease these computations, we rescale the triangle to have a height of $1$.  {We also introduce temporary, local polar coordinates $(r, \theta)$, which notably are distinct from the cylindrical polar coordinates used for the full sarcomere, and these have been added to Fig.~\ref{Figure 8}.} Assuming that  the pressure gradient is entirely axial (i.e.~along $z$), and varies over length scales much greater than the size of the triangle, we calculate the flow $w$ driven by a uniform axial dimensionless pressure gradient of 1 by solving the resultant dimensionless Stokes equation in the triangle
\begin{equation}
\nabla^2 w = \frac{\partial p}{\partial z} = 1.
\label{eq:w}
\end{equation}
We next apply a substitution $w^* = w - r^2/4$ and solve the resultant Laplace equation in plane polar coordinates
\begin{equation}
\frac{\partial^2 w^*}{\partial^2 r} + \frac{1}{r}\frac{\partial w^*}{\partial r} + \frac{1}{r^2}\frac{\partial^2 w^*}{\partial^2 \theta} = 0.
\end{equation}

We now assume that the length scale on which the applied pressure gradient $\partial P/\partial z$ changes is much larger than the size of each triangle. Consequently, the axial flow profile in each triangle and the three adjacent triangles is identical, after appropriate reflections. Therefore, symmetry conditions manifest along each edge of the triangle, $\partial w/\partial n = 0$. Note also the no-slip condition $w = 0$ on the surfaces of the filaments. Thus we obtain a boundary condition along each boundary, and the problem is fully determined with a unique solution.

The solution to Eq.~\eqref{eq:w} for the axial flow that satisfies the symmetry conditions at $\theta = 0$ and $\theta = \pi/6$ is given by the truncated series
\begin{equation}
w = A + B\log(r) + \frac{r^2}{4} + \sum_{k=1}^N \left(C_k r^{6k} + D_k r^{-6k}\right)\cos(6k\theta).
\end{equation}
Once again, this value of $N$ is temporary and distinct from the value of $N$ seen in Eq. \eqref{eq:pressure_truncated}.  {We use $R_{thick}$ to denote the rescaled radius of the thick myosin filament.} Within the thick and overlap regions, applying $w = 0$ on the thick filament gives
\begin{equation}
\begin{split}
w = &B\log\left(\frac{r}{R_{thick}}\right) + \frac{r^2 - R_{thick}^2}{4} \\
&+ \sum_{k=1}^N C_kr^{6k}\left(1 - \left(\frac{r}{R_{thick}}\right)^{-6k}\right)\cos(6k\theta),
\end{split}
\end{equation}
whilst in the thin region, regularity at $r = 0$ gives simply
\begin{equation}
w = A + \frac{r^2}{4} + \sum_{k=1}^N C_k r^{6k}\cos(6k\theta).
\label{eq:wA}
\end{equation}
Each of these solutions involves $N+1$ coefficients, which we determine numerically by applying the unused symmetry condition on the lower edge, and the no-slip conditions on the thin blue filaments (if present) via a least squares method. Upon computing these coefficients, we have thus obtained a semi-analytic solution for $w$.

Noting that the height of the triangle was rescaled to $1$, giving an area of $1/2\sqrt{3}$, and recalling that permeability is defined as the ratio between fluid flux per unit area and pressure gradient, we then calculate a dimensionless permeability in each region $k_\|^{(j)} = 2\sqrt{3}\mathcal{F}^{(j)}$, where $\mathcal{F}^{(j)} = \int w^{(j)} ~ dA^{(j)}$ is the total fluid flux through the triangle.  {The resulting values for the dimensionless axial permeabilities of real sarcomeres ($k_\|^{(j)},\, j=1,2,3$) are given in Table~\ref{table2} in the main text.  {These permeabilities must be re-dimensionalised to have the appropriate dimensions of length squared by multiplying $k_\|^{(j)}$ by $H^2$, where $H = D_{10}/\sqrt{3}$ is the original, dimensional height of the triangle. To then incorporate these dimensional permeabilities into Eq.~\eqref{eq:pressure_truncated}, they must be non-dimensionalised under the same scheme as the sarcomere. Since each $k_\|^{(j)}$ refers to axial flow, the appropriate scaling involves a division by $L^2$, where $L$ is the dimensional sarcomere length. Therefore, each $k_\|^{(j)}$ in Table~\ref{table2} should overall by multiplied by ${D_{10}^2}/{3L^2}$ to be used in Eq.~\eqref{eq:pressure_truncated}.}}

 {It should be noted that the least squares method is typically more numerically robust if we rescale the terms of the semi-analytic solution to not increase rapidly with $n$. For example, since  the maximum possible value of $r$ is $2/\sqrt{3}$, we may rewrite Eq.~\eqref{eq:wA} as
\begin{equation}
w = A + \frac{r^2}{4} + \sum_{k=1}^N \tilde{C}_k \left(\frac{\sqrt{3}}{2}r\right)^{6k}\cos(6k\theta).
\end{equation}
Here, the value of $N$ required to produce accurate results, as given in Table~\ref{table2}, depends on the packing ratio and varies between the regions of the sarcomere, as well as the particular implementation of the least squares matching. In general, retaining a few dozen terms allows for excellent accuracy, such that the values given in Table~\ref{table2} are accurate to the given number of digits.}

\section{Calculating the traction parameter $\beta$} 

\label{appendix:beta}

With the axial permeabilities known, there remains the task of calculating the value of the traction parameter $\beta$. This may be done by adapting the axial flow solution from Appendix~\ref{appendix:axial_permeabilities} in the overlap region (since $\beta$ exists only in the overlap region), by removing the pressure gradient, and replacing the boundary conditions on the thin blue filament(s) with $w = 1$. The semi-analytic solution that satisfies $w = 0$ on the thick filament $r = R_{thick}$ and the symmetry conditions $\partial w/\partial \theta = 0$ on the vertical edge $\theta = 0$ and the hypotenuse $\theta = \pi/3$ is given by
\begin{equation}
w = B\log\left(\frac{r}{R_{thick}}\right) + \sum_{k=1}^N C_kr^{6k}\left(1 - \left(\frac{r}{R_{thick}}\right)^{-6k}\right)\cos(6k\theta).
\end{equation}
Again, the coefficients are determined numerically by a least squares method, matching the remaining symmetry condition along the lower edge, as well as the no-slip condition $w = 1$ on the thin blue filament(s). We then calculate the value of $\beta$ from the flux, $\mathcal{F} = \int w ~ dA$, as
\begin{equation}
\beta = \frac{2\sqrt{3}}{\Phi^{(2)}}\mathcal{F}.
\end{equation}
Note that $\beta$ requires no re-dimensionalisation. The resulting values for $\beta$ of real sarcomeres are given in Table~\ref{table2} in the main text.

\section{Lateral permeabilities}
\label{appendix:lateral_permeabilities}

The lateral permeabilities are more complicated to determine, but we can adapt a semi-analytic scheme, originally used by Sangani \& Acrivos~\cite{Sangani_Acrivos_1982}, who considered a hexagonal array of cylinders of equal radii (i.e.~our thick region) as a series of repeating rectangles with quarter-circles in a pair of opposite vertices. We could immediately apply this method to the thick region, but in the other regions, the thin filaments would render the method ineffective. However,  by considering additional symmetries not originally noted in Ref.~\cite{Sangani_Acrivos_1982}, we can significantly reduce the size of the domain and the number of numerical boundary conditions, as well as generalise the problem to the presence or absence of thick and/or thin filaments. 

To do so, we simultaneously solve for the two solutions resulting from two orthogonal pressure gradients on the same triangles illustrated in Fig.~\ref{Figure 2}F-H,  {and reproduced for a packing ratio of 5 in Fig.~\ref{Figure 8},} thereby reducing the size of the domain on which the solution is solved by a factor of six compared to the original study~\cite{Sangani_Acrivos_1982}. Within the thick region, there is only one boundary condition that we must match numerically, compared to three originally~\cite{Sangani_Acrivos_1982}. Finally, we can include thin filaments in any of the three packing ratios of Fig.~\ref{Figure 2}C-H, and even remove the thick filament, and can still solve the problem semi-analytically.

As with the axial permeability calculations, we rescale the triangle from a dimensional height of $H { = D_{10}/\sqrt{3}}$ to a dimensionless height of $1$, and we employ a temporary, local polar coordinate system{, $(r, \theta)$, as in Fig.~\ref{Figure 8}}. We begin by identifying the biharmonic equation for the streamfunction $\psi$, relating to the vorticity $\omega$ 
\begin{equation}
\nabla^4\psi = -\nabla^2\omega = 0,
\end{equation}
which produces a velocity
\begin{equation}
\mathbf{u} = \frac{1}{r}\frac{\partial \psi}{\partial \theta}\mathbf{e}_r - \frac{\partial \psi}{\partial r}\mathbf{e}_\theta.
\end{equation}
On the surfaces of filaments, we have no slip, giving zero normal derivative, $\partial \psi / \partial n = 0$. We also have zero tangential derivative, $\partial \psi / \partial t = 0$, which we implement as $\psi = const$, where the value of the constant is set by a flux condition. It will be most convenient to universally specify the Darcy flux (i.e.~flux per unit area) to be $1$, so the actual flux through the triangle will depend on the direction of the pressure gradient. Specifically, we set the flux across the triangle to be $1$ for a horizontal pressure gradient, and $1/\sqrt{3}$ for a vertical pressure gradient. We then consider four different directions for the pressure gradient - one horizontal, one vertical, one normal to the hypotenuse, and one parallel to the hypotenuse, with streamfunctions $\psi_h, \psi_v, \psi_\perp$ and $\psi_\|$ respectively. As with the axial case, we assume that changes in $\nabla P$ occur over length scales much larger than the size of each triangle, giving symmetry conditions along the edges of the triangle, dependent on the direction of the applied pressure gradient. Whenever the pressure gradient is parallel to a side of the triangle, we have $\partial \psi / \partial t = \omega = 0$ at that side boundary.  When the pressure gradient is normal to a side of the triangle, we have $\partial \psi / \partial n = \partial \omega / \partial n = 0$ there. We relate the four solutions by
\begin{align}
\psi_\| &= \frac{\sqrt{3}}{2}\psi_v - \frac{1}{2}\psi_h, \\
\psi_\perp &= \frac{1}{2}\psi_v + \frac{\sqrt{3}}{2}\psi_h.
\end{align}
We apply boundary conditions on the hypotenuse for $\psi_\|$ and $\psi_\perp$, as described above. Therefore, we have four boundary conditions along the hypotenuse for $\psi_v$ and $\psi_h$ together, leading to  four conditions on each and every boundary for the coupled system $(\psi_v, \psi_h)$, which is hence fully determined.

Complete expressions for the streamfunctions, for the three regions of the sarcomere, with corresponding expressions for vorticity and pressure, are provided in the following subsections. Note that, for any given pressure gradient, giving streamfunction $\psi$ in the triangles of Fig.~\ref{Figure 2}F-H, we can use appropriate superpositions of $\psi_h$ and $\psi_v$ (along with additive constants) to determine the extension of the streamfunction to the other triangles, and hence the entire hexagonal cell. Somewhat remarkably, we find that the solution in the entire hexagonal cell is simply $\psi$ extended to a domain of all $\theta$ between $0$ and $2\pi$. 

Finally, the expressions for pressure allow us to determine the dimensionless pressure change $\Delta p$ over a length $l$ through the hexagonal cell, giving each permeability as
\begin{equation}
k_\perp^{(j)} = -\frac{l}{\Delta p^{(j)}}.
\end{equation}
As mentioned in the main text,  the value of $k_\perp^{(j)}$ is independent of the chosen direction for the applied pressure gradient. The resulting values for the lateral permeabilities of real sarcomeres ($k_\perp^{(j)},\, j=1,2,3$) are listed in Table~\ref{table2} in the main text.  {As with the axial permeabilities, each $k_\perp^{(j)}$ must be multiplied by ${D_{10}^2}/{3}$ to be re-dimensionalised. Since each $k_\perp^{(j)}$ refers to lateral flow, they must then be divided by $R^2$, where $R$ is the dimensional sarcomere radius, in order to have the same non-dimensionalisation used for the sarcomere. This means the $k_\perp^{(j)}$ in Table~\ref{table2} must overall be multiplied by ${D_{10}^2}/{3R^2}$ to be used in Eq.~\eqref{eq:pressure_truncated}.}

\subsection{Solutions in the thick and overlap regions}
We obtain solutions for the streamfunctions for horizontal and vertical flow:
\begin{equation}
\begin{split}
\psi_h &= A_0r^3\left(1 - 2R_{thick}^2r^{-2} + R_{thick}^4r^{-4}\right)\cos(\theta) \\
&+ B_0r\left(2\log\left(r\right) - 2\log(R_{thick}) - 1 + R_{thick}^2r^{-2}\right)\cos(\theta) \\
+ \sum_{n=1}^\infty &A_nr^{6n+3}\left((6n+1) - (6n+2)R_{thick}^2r^{-2} + R_{thick}^{12n+4}r^{-(12n+4)}\right)\cos\left((6n+1)\theta\right) \\
&+ B_nr^{6n+1}\left(1 - (6n+1)R_{thick}^{12n}r^{-12n} + 6nR_{thick}^{12n+2}r^{-(12n+2)}\right)\cos\left((6n+1)\theta\right) \\
&+ C_nr^{6n+1}\left((6n-1) - 6nR_{thick}^2r^{-2} + R_{thick}^{12n}r^{-12n}\right)\cos\left((6n-1)\theta\right) \\
&+ D_nr^{6n-1}\left(1 - (6n-1)R_{thick}^{12n-4}r^{-(12n - 4)} + (6n-2)R_{thick}^{12n-2}r^{-(12n-2)}\right)\cos\left((6n-1)\theta\right),
\end{split}
\end{equation}
and
\begin{equation}
\begin{split}
\psi_v &= A_0r^3\left(1 - 2R_{thick}^2r^{-2} + R_{thick}^4r^{-4}\right)\sin(\theta) \\
&+ B_0r\left(2\log(r) - 2\log(R_{thick}) - 1 + R_{thick}^2r^{-2}\right)\sin(\theta) \\
+ \sum_{n=1}^\infty &A_nr^{6n+3}\left((6n+1) - (6n+2)R_{thick}^2r^{-2} + R_{thick}^{12n+4}r^{-(12n+4)}\right)\sin\left((6n+1)\theta\right) \\
&+ B_nr^{6n+1}\left(1 - (6n+1)R_{thick}^{12n}r^{-12n} + 6nR_{thick}^{12n+2}r^{-(12n+2)}\right)\sin\left((6n+1)\theta\right) \\
&- C_nr^{6n+1}\left((6n-1) - 6nR_{thick}^2r^{-2} + R_{thick}^{12n}r^{-12n}\right)\sin\left((6n-1)\theta\right) \\
&- D_nr^{6n-1}\left(1 - (6n-1)R_{thick}^{12n-4}r^{-(12n - 4)} + (6n-2)R_{thick}^{12n-2}r^{-(12n-2)}\right)\sin\left((6n-1)\theta\right).
\end{split}
\end{equation}
These give vorticities
\begin{equation}
\begin{split}
-\frac{1}{4} &\omega_h = \left(2A_0r + B_0r^{-1}\right)\cos(\theta) \\
&+ \sum_{n=1}^\infty (6n+1)\left((6n+2)A_nr^{6n+1} + 6nR_{thick}^{12n}B_nr^{-(6n+1)}\right)\cos((6n+1)\theta) \\
&+ \sum_{n=1}^\infty (6n-1)\left(6nC_nr^{6n-1} + (6n-2)R_{thick}^{12n-4}D_nr^{-(6n-1)}\right)\cos((6n-1)\theta),
\end{split}
\end{equation}
and
\begin{equation}
\begin{split}
-\frac{1}{4} &\omega_v = \left(2A_0r + B_0r^{-1}\right)\sin(\theta) \\
&+ \sum_{n=1}^\infty (6n+1)\left((6n+2)A_nr^{6n+1} + 6nR_{thick}^{12n}B_nr^{-(6n+1)}\right)\sin((6n+1)\theta) \\
&- \sum_{n=1}^\infty (6n-1)\left(6nC_nr^{6n-1} + (6n-2)R_{thick}^{12n-4}D_nr^{-(6n-1)}\right)\sin((6n-1)\theta).
\end{split}
\end{equation}
We calculate dimensionless pressure analytically by noting that $\partial p/\partial \theta = r\partial \omega/\partial r$:
\begin{equation}
\begin{split}
-\frac{1}{4} &p_h = p_0 + \left(2A_0r - B_0r^{-1}\right)\sin(\theta) \\
&+ \sum_{n=1}^\infty (6n+1)\left((6n+2)A_nr^{6n+1} - 6nR_{thick}^{12n}B_nr^{-(6n+1)}\right)\sin((6n+1)\theta) \\
&+ \sum_{n=1}^\infty (6n-1)\left(6nC_nr^{6n-1} - (6n-2)R_{thick}^{12n-4}D_nr^{-(6n-1)}\right)\sin((6n-1)\theta),
\end{split}
\end{equation}
and
\begin{equation}
\begin{split}
-\frac{1}{4} &p_v = p_0 + \left(-2A_0r + B_0r^{-1}\right)\cos(\theta) \\
&+ \sum_{n=1}^\infty (6n+1)\left(-(6n+2)A_nr^{6n+1} + 6nR_{thick}^{12n}B_nr^{-(6n+1)}\right)\cos((6n+1)\theta) \\
&- \sum_{n=1}^\infty (6n-1)\left(-6nC_nr^{6n-1} + (6n-2)R_{thick}^{12n-4}D_nr^{-(6n-1)}\right)\cos((6n-1)\theta).
\end{split}
\end{equation}

\subsection{Solutions in the thin region}
In the thin region, the solutions are different due to the regularity condition at $r = 0$. We have
\begin{equation}
\begin{split}
\psi_h = A_0r^3\cos(\theta) + B_0r\cos(\theta) &+ \sum_{n=1}^\infty \left(A_nr^{6n+3} + B_nr^{6n+1}\right)\cos\left((6n+1)\theta\right) \\
&+ \sum_{n=1}^\infty \left(C_nr^{6n+1} + D_nr^{6n-1}\right)\cos\left((6n-1)\theta\right),
\end{split}
\end{equation}
\begin{equation}
\begin{split}
\psi_v = A_0r^3\sin(\theta) + B_0r\sin(\theta) &+ \sum_{n=1}^\infty \left(A_nr^{6n+3} + B_nr^{6n+1}\right)\sin\left((6n+1)\theta\right) \\
&- \sum_{n=1}^\infty \left(C_nr^{6n+1} + D_nr^{6n-1}\right)\sin\left((6n-1)\theta\right).
\end{split}
\end{equation}
These give vorticities
\begin{align}
-\frac{1}{4}\omega_h &= 2A_0r\cos(\theta) + \sum_{n=1}^\infty (6n+2)A_nr^{6n+1}\cos((6n+1)\theta) + 6nC_nr^{6n-1}\cos((6n-1)\theta), \\
-\frac{1}{4}\omega_v &= 2A_0r\sin(\theta) + \sum_{n=1}^\infty (6n+2)A_nr^{6n+1}\sin((6n+1)\theta) - 6nC_nr^{6n-1}\sin((6n-1)\theta),
\end{align}
and dimensionless pressures
\begin{align}
-\frac{1}{4}p_h &= p_0 + 2A_0r\sin(\theta) + \sum_{n=1}^\infty (6n+2)A_nr^{6n+1}\sin((6n+1)\theta) + 6nC_nr^{6n-1}\sin((6n-1)\theta), \\
-\frac{1}{4}p_v &= p_0 - 2A_0r\cos(\theta) + \sum_{n=1}^\infty -(6n+2)A_nr^{6n+1}\cos((6n+1)\theta) + 6nC_nr^{6n-1}\cos((6n-1)\theta).
\end{align}

\section{Boundary and interface conditions}
\label{appendix:boundary_and_interface_conditions}
The solution for the Darcy pressure is the truncated series of Eq.~\eqref{eq:pressure_truncated}:
\begin{equation}
P^{(j)}(r,z) = \sum_{n=1}^N \left[A_{n}^{(j)}e^{G_{n}^{(j)}z} + B_{n}^{(j)}e^{-G_{n}^{(j)}z}\right]J_0(\lambda_{n} r),
\end{equation}
where $J_0$ is the zero order Bessel function of the first kind, $\lambda_n$ is the $n$th zero of $J_0$, and $G_{n}^{(j)} = \lambda_n \sqrt{k_{\perp}^{(j)}/k_{\|}^{(j)}}$. To identify the values of the coefficients, we apply the boundary and interface conditions \eqref{bc2}~-~\eqref{bc7} via the following orthogonality condition for Bessel functions~{\cite{Abramowitz_Stegun_1970}}:
\begin{equation}
\int_0^1 rJ_0(\lambda_n r)J_0(\lambda_m r) dr = \frac{1}{2}\delta_{m,n}J_1(\lambda_n)^2.
\end{equation}
For each $n$, we define
\begin{equation}
I_n = \frac{2\int_0^1 rJ_0(\lambda_n r) dr}{J_1(\lambda_n)^2}  { = \frac{2}{\lambda_n J_1(\lambda_n)}},
\end{equation}
{where we have evaluated the integral using the definition of $J_0(x)$
\begin{equation}
x^2J_0''(x) + xJ_0'(x) + x^2J_0(x) = 0,
\end{equation}
and the fact that $J_0'(x) = -J_1(x)$.} We hence obtain the following equation for the coefficients, independently for each value of $n$,
\begin{equation}
\mathbf{M}\left(\begin{matrix}A_{1,n} \\ B_{1,n} \\ A_{2,n} \\ B_{2,n} \\ A_{3,n} \\ B_{3,n}\end{matrix}\right) = \left(\begin{matrix}0 \\ 0 \\ \left(\beta\Phi_2 + \left(1 - \Phi_3\right)\right)I_n \\ \left(\Phi_3 - \beta\Phi_2\right)I_n \\ 0 \\ 0\end{matrix}\right),
\label{eq:M1}
\end{equation}
where
\begin{equation}
\mathbf{M} = \left(\begin{matrix}1 & -1 & 0 & 0 & 0 & 0 \\ 0 & 0 & 0 & 0 & e^{G_{n}^{(3)}} & -e^{-G_{n}^{(3)}} \\ k_{\| 1}G_{n}^{(1)}e^{G_{n}^{(1)}L_1} & -k_{\|1}G_{n}^{(1)} e^{-G_{n}^{(1)}L_1} & -k_{\|2}G_{n}^{(2)}e^{G_{n}^{(2)}L_1} & k_{\|2}G_{n}^{(2)}e^{-G_{n}^{(2)}L_1} & 0 & 0 \\ 0 & 0 & k_{\| 2}G_{n}^{(2)}e^{G_{n}^{(2)}L_2} & -k_{\|2}G_{n}^{(2)} e^{-G_{n}^{(2)}L_2} & -k_{\|3}G_{n}^{(3)}e^{G_{n}^{(3)}L_2} & k_{\|3}G_{n}^{(3)}e^{-G_{n}^{(3)}L_2} \\ e^{G_{n}^{(1)}L_1} & e^{-G_{n}^{(1)}L_1} & -e^{G_{n}^{(2)}L_1} & -e^{-G_{n}^{(2)}L_1} & 0 & 0 \\ 0 & 0 & e^{G_{n}^{(2)}L_2} & e^{-G_{n}^{(2)}L_2} & -e^{G_{n}^{(3)}L_2} & -e^{-G_{n}^{(3)}L_2}\end{matrix}\right).
\label{eq:M2}
\end{equation}
Provided with values for the permeabilities, we solve these equations numerically for each $n$, thereby obtaining the coefficients $A_n^{(j)}$ and $B_n^{(j)}$ and hence the pressure $P$. However, it is generally more numerically robust to write this system as
\begin{equation}
P^{(j)}(r,z) = \sum_{n=1}^N \left[\tilde{A}_{n}^{(j)}e^{G_{n}^{(j)}(z-L_j)} + \tilde{B}_{n}^{(j)}e^{-G_{n}^{(j)}(z-L_{j-1})}\right]J_0(\lambda_{n} r),
\end{equation}
where $L_0 = 0$ and $L_3 = 1$, with corresponding matrix equation
\begin{equation}
\tilde{\mathbf{M}}\left(\begin{matrix}\tilde{A}_{1,n} \\ \tilde{B}_{1,n} \\ \tilde{A}_{2,n} \\ \tilde{B}_{2,n} \\ \tilde{A}_{3,n} \\ \tilde{B}_{3,n}\end{matrix}\right) = \left(\begin{matrix}0 \\ 0 \\ \left(\beta\Phi_2 + \left(1 - \Phi_3\right)\right)I_n \\ \left(\Phi_3 - \beta\Phi_2\right)I_n \\ 0 \\ 0\end{matrix}\right),
\end{equation}
where the matrix $\tilde{\mathbf{M}}$ is given by
\begin{equation}
\tilde{\mathbf{M}} = \left(\begin{matrix}e^{-G_{n}^{(1)}L_1} & -1 & 0 & 0 & 0 & 0 \\ 0 & 0 & 0 & 0 & 1 & -e^{-G_{n}^{(3)}(1-L_2)} \\ k_{\| 1}G_{n}^{(1)} & -k_{\|1}G_{n}^{(1)} e^{-G_{n}^{(1)}L_1} & -k_{\|2}G_{n}^{(2)}e^{-G_{n}^{(2)}(L_2-L_1)} & k_{\|2}G_{n}^{(2)} & 0 & 0 \\ 0 & 0 & k_{\| 2}G_{n}^{(2)} & -k_{\|2}G_{n}^{(2)} e^{-G_{n}^{(2)}(L_2-L_1)} & -k_{\|3}G_{n}^{(3)}e^{-G_{n}^{(3)}(1-L_2)} & k_{\|3}G_{n}^{(3)} \\ 1 & e^{-G_{n}^{(1)}L_1} & -e^{-G_{n}^{(2)}(L_2-L_1)} & -1 & 0 & 0 \\ 0 & 0 & 1 & e^{-G_{n}^{(2)}(L_2-L_1)} & -e^{-G_{n}^{(3)}(1-L_2)} & -1\end{matrix}\right).
\end{equation}
Under these rescalings, both $P$ and $\tilde{\mathbf{M}}$ contain only terms that are at most $\mathcal{O}(1)$ (or more accurately, $\mathcal{O}(n)$), greatly improving numerical accuracy. In practice, we find that the coefficients $\tilde{A}_n^{(3)}$ and $\tilde{B}_n^{(1)}$ decay exponentially with $n$, whilst the remaining coefficients decay like $n^{-3/2}$. We find that, upon reaching $n = 50$, all coefficients have decayed to less than $2\%$ of that coefficient's peak value across all values of $n$, and the decay is usually significantly greater than this, around $0.1\%$. From this, we conclude that a truncation of $N = 50$ is sufficient to produce accurate values for the pressure field, and hence the velocity fields.

\section{FEM computations}
\label{appendix:FEM_computations}
The full computations (finite element method) were performed dimensionally within COMSOL Multiphysics Version 5.6~\cite{COMSOL}, 
using a stationary study and creeping flow physics interface. We used COMSOL's option for a normal mesh size, with a maximum element size of $47.4~$nm, maximum element growth rate of 1.15, curvature factor of 0.6 and resolution of narrow regions of 0.7. However we manually reduced the minimum element size to $6.5~$nm, allowing filament cross-sections to be approximated as octagons by the mesh. We applied no-slip conditions on the surfaces of filaments and disks, and symmetry conditions at $z = 0$ when the M disk was absent. The Z disk and thin filaments had a contraction speed of $1000~$nm/s, and the dynamic viscosity of the fluid was set to that of water at room-temperature, $0.001~$Pa s, though both of these are largely irrelevant due to linearity. Symmetry conditions allowed us to consider only a $30^\circ$ sector of the sarcomere. We approximated a stress-free far-field condition by extending the Z disk and M disk boundaries (whether present or absent), and their corresponding boundary conditions, to twice the sarcomere radius, effectively producing a sarcomere with twice the radius, but with filaments present within only the inner half-radius. We then applied a stress-free condition on the enlarged outer radial boundary.

{\section{Advection-reaction-diffusion}
\label{appendix:ARD}

\begin{figure}
\centering
\includegraphics[width=0.72\textwidth]{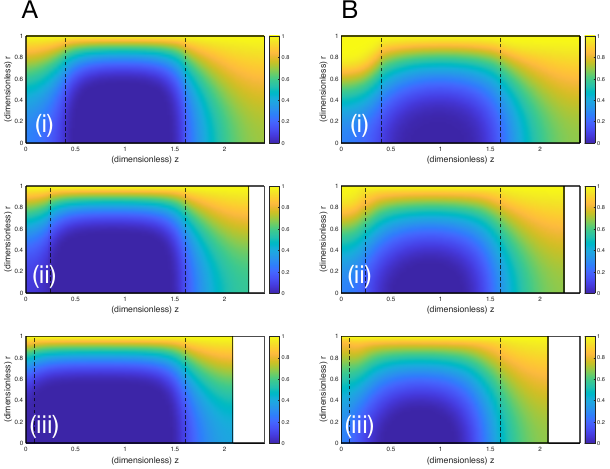}
\caption{{ATP concentration over the course of a contraction, in a human cardiac sarcomere with no advection (A) and advection  by the fluid flow (B). Snapshots of ATP concentration are shown at low (i), intermediate (ii) and high (iii) levels of human cardiac sarcomere contraction. Horizontal axis represents the dimensionless axial coordinate of the sarcomere, and the vertical axis represents the dimensionless radial coordinate. Dashed lines indicate transitions between the key regions of the sarcomere, and reaction occurs in the middle region only}.}
\label{Figure 9}
\end{figure}

Experiments have shown that diffusion of ATP inside the sarcomere is surprisingly slow, with diffusive coefficient estimated to be $D \approx 2 \times 10^{-15}~$m$^2$/s, several orders of magnitude lower than what would be observed in water~\cite{Alekseev_Guzun_Reyes_Pison_Schlattner_Selivanov_Cascante_2016}. This is likely due to the dense ecology of substrates and enzymes within the sarcoplasm~\cite{Alekseev_Guzun_Reyes_Pison_Schlattner_Selivanov_Cascante_2016, Hargreaves_Spriet_2020}. For contraction consistent with human cardiac sarcomeres, this gives a relevant P\'eclet number of $Pe \approx 125$, suggesting that advection has significant impact on ATP distribution~\cite{Rodriguez_Hunter_Royce_Leppo_Douglas_Weisman_1992, Stenger_Spiro_1961, Shimomura_Iwamoto_Doan_Ishiwata_Sato_Suzuki_2016, ter_Keurs_Diao_Deis_2010, Vinnakota_Bassingthwaighte_2004, Burbaum_Schneider_Scholze_Böttcher_Baumeister_Schwille_Plitzko_Jasnin_2021, Moore_Ruska_1957, Linke_Popov_Pollack_1994, Oda_Yanagisawa_2020}. We model the reaction of ATP as occurring in the overlap region only, and during contraction only. We quantify the reaction using Michaelis-Menten kinetics, with reaction rate $V_m C/ (K_m + C)$, where $C$ is the ATP concentration, maintained at $10~$mM outside the sarcomere, and Michaelis constants $V_m = 1~$mM and $K_m = 0.01~$mM~\cite{Alekseev_Guzun_Reyes_Pison_Schlattner_Selivanov_Cascante_2016}. We non-dimensionalised the system, scaling all lengths with sarcomere radius $R$, and scaling ATP concentration with the external ATP concentration of $10~$mM. We set boundary conditions $C = 1$ at $r = 1$, $\partial C/\partial r = 0$ at $r = 0$ and $\partial C/\partial z = 0$ at $z = 0$ and $z = L$, where $L$ is the dimensionless sarcomere length, and solved for the evolution of the system using an advective grid method that stretches with the sarcomere as it contracts and relaxes sinusoidally, with continuity of both $C$ and substrate flux applied at interfaces between regions, until a periodic state was reached. Comparisons between this periodic state, and the state achieved in the absence of fluid advection, are shown in Fig.~\ref{Figure 9} (A: no flow;   B: advection by flow included) at various stages of contraction (i, ii, iii). We clearly see that the fluid flow significantly reduces the sizes of ATP ``deadzones'', i.e.~regions of very low ATP concentration, and overall increases the average reaction rate by approximately $26 \%$, representing a significant improvement in sarcomere activity. These preliminary calculations warrant further investigations.}

{\section{Varying permeabilities in the thin region}
\label{appendix:thin_region}

\begin{figure}
\centering
\includegraphics[width = \textwidth]{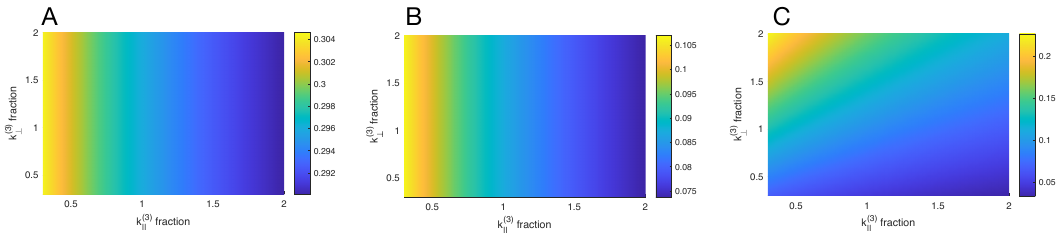}
\caption{{Total dimensionless radial efflux in regions 1 (A), 2 (B) and 3(C) due to variations of $k^{(3)}_\|$ and $k^{(3)}_\perp$ by factors between $0.3$ and $2$, compared to their values in Table~\ref{table2}. The porosity $\Phi^{(3)}$ in the thin region is fixed at its established value used in previous examples in the main text.}}
\label{Figure 10}
\end{figure}

In mammalian sarcomeres, the presence of the giant protein titin and the disorganisation of the thin filaments within the thin region may give pause for concern regarding the calculated permeabilities $k^{(3)}_\|$ and $k^{(3)}_\perp$. As discussed in the main text, a semi-analytic calculation of these permeabilities would be unfeasible, however they may be calculated numerically (provided sufficiently accurate physical models of the molecular microstructure are available) or possibly measured experimentally. To demonstrate the potential for the Darcy model to be applied to this new system, we plot in Fig.~\ref{Figure 10} the total radial efflux leaving the sarcomere, in each of the three regions (A, B, C for regions 1, 2, 3 respectively), for the same sarcomere as in Fig.~\ref{Figure 7}c, consistent with mammalian sarcomeres, with the only changes being variation of $k^{(3)}_\|$ and $k^{(3)}_\perp$ from their standard values in Table \ref{table2} by dimensionless factors ranging between $0.3$ (making them comparable to the permeabilities in the overlap region) and $2$. These results demonstrate how the Darcy model can be applied to a continuous range of parameters (as well as highlighting the independence of the radial efflux in the thick and overlap regions on the lateral permeability in the thin region, $k^{(3)}_\perp$). Importantly, the Darcy model can be applied to any system with specified, even arbitrary permeabilities, without the need to resolve the precise physical structure of the filaments.}

\bibstyle{unsrt}
\bibliography{bibliography_sarcomere}
\end{document}